\newcommand{\ignore}[1]{}

\documentclass[9pt, conference]{IEEEtran}
\usepackage{url}
\usepackage{multirow}
\usepackage{subfigure}
\usepackage{epsfig}
\usepackage{graphicx}
\usepackage{amsmath}
\usepackage{amssymb}
\usepackage{color}
\usepackage{setspace}
\usepackage{cite}
\usepackage{hhline}   
\usepackage{algorithm}
\usepackage{algpseudocode}
\usepackage{flushend}
\usepackage{url}
\usepackage{hyperref}
\usepackage[normalem]{ulem}
\usepackage{graphics}
\usepackage{tikz}
\usepackage{array}
\usepackage{animate}
\usepackage{titlesec}
\usepackage{balance}
\usepackage{enumitem}
\usepackage{booktabs}
\usepackage{pifont}
\usepackage{balance}

\titlespacing*{\section}{0pt}{3pt}{3pt}
\titlespacing*{\subsection}{0pt}{3pt}{2pt}
\titlespacing*{\subsubsection}{0pt}{3pt}{2pt}

\begin{document}

\title{CarbonPATH: Carbon-aware pathfinding and architecture optimization for chiplet-based AI systems}
\vspace{-1mm}
\author{Chetan Choppali Sudarshan, Jiajun Hu, Aman Arora, and Vidya A. Chhabria \\ Arizona State University}


\maketitle
\begin{abstract}

\noindent 
The exponential growth of AI has created unprecedented demand for computational resources, pushing chip designs to the limit while simultaneously escalating the environmental footprint of computing. As the industry transitions toward heterogeneous integration (HI) to address the yield and cost challenges of monolithic scaling, minimizing the carbon cost of these complex HI systems becomes critical. To fully exploit HI, a co-design approach spanning application, architecture, chip, and packaging is essential. However, this creates a vast design space with competing objectives, specifically the trade-offs between performance, cost, and carbon footprint (CFP) for sustainability. CarbonPATH is an early-stage pathfinding framework designed to address this multi-objective challenge. It identifies optimized HI systems by co-designing workload mapping, architectural parameters, and packaging technologies, while treating sustainability as a first-class design constraint. The framework accounts for a wide range of factors, including compute and memory sizes, chiplet technology nodes, communication protocols, integration style (2D, 2.5D, 3D), operational CFP, embodied CFP, and interconnect type. Using simulated annealing, CarbonPATH explores this high-dimensional space to identify solutions that balance traditional metrics against environmental impact.
By capturing interactions across applications, architectures, chiplets, and packaging, CarbonPATH uncovers system-level solutions that traditional methods often miss due to restrictive assumptions or limited scope.

\end{abstract}

\section{Introduction}
\label{sec:intro}

\noindent
{\bf Motivation and problem.}
As AI systems scale to unprecedented sizes, designing hardware that balances performance with environmental sustainability is becoming a fundamental challenge. These models require executing trillions of operations and moving terabytes of data, which far exceeds the capability of a single accelerator. The most performance-efficient solution would be a monolithic chip that accommodates all model parameters and intermediate activations on a single die~\cite{peng2023chiplet}. However, the slowdown of Moore’s law and Dennard scaling, combined with reticle size limits and escalating costs, makes this approach impractical and unsustainable~\cite{hi-1, eco-chip}. The manufacturing of these massive monolithic dies increases environmental impact, as lower die yields result in a larger manufacturing carbon footprint (CFP) and increased silicon waste~\cite {hi-1,eco-chip}.
Advanced packaging and heterogeneous integration (HI) have therefore emerged as promising alternatives~\cite{eda-hi-survey-dac25}. By assembling multiple specialized dies within a single package, designers can approximate the performance of a large monolithic accelerator while improving yield, reducing non-recurring engineering costs, reusing IP across product generations~\cite{chiplet-costonly}, and lowering CFP~\cite{eco-chip}.


The design of chiplet-based AI accelerators, however, presents significant challenges. System-level performance, power or energy efficiency, area, cost (PPAC), and CFP depend not only on application and architectural choices, such as resource allocation, workload mapping, on-chip memory sizes, sizing of compute, and dataflow, but also on chip and packaging decisions, including placement, interconnect topology, and the choice between 2.5D and 3D integration technologies. 
These choices are highly interdependent. For example, modifying resource allocation affects bandwidth requirements, which in turn constrains packaging choices and impacts overall system cost and CFP. Most existing research often addresses architecture and packaging independently~\cite{peng2023chiplet,wang2022application,coskun2022cross,lin2020deep,shao2019simba,tan2021nn,li2021scaling}, which leads to suboptimal designs for HI systems. 


\begin{figure}[t]
\centering
\includegraphics[width=0.75\linewidth]{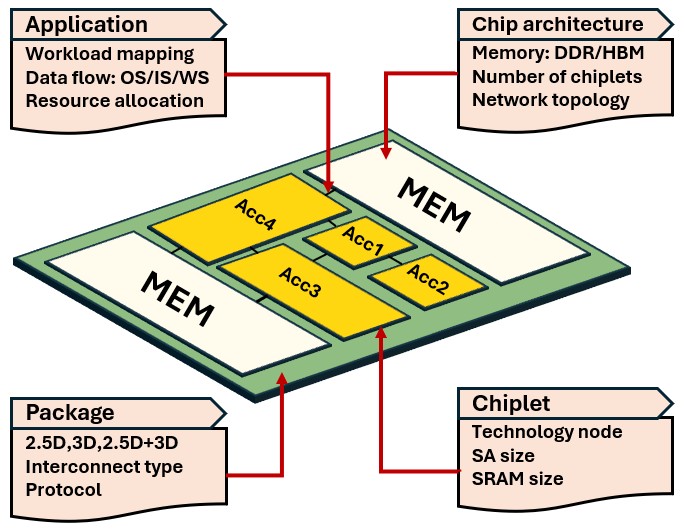}
\caption{System-level co-design across application, chip architecture, chiplet configuration, and package choices for chiplet-based accelerators.}
\label{fig:co-design-compute-stack}
\end{figure}

Further, as major technology firms increasingly commit to strict decarbonization timelines~\cite{apple-decarbonization-goal, google-decarbonization-goal, meta-decarbonization-goal}, minimizing total CFP has moved from a corporate initiative to a fundamental engineering constraint. Total CFP comprises embodied emissions from the manufacturing of dies and packaging of the HI system, and operational emissions from active workload execution. While HI offers a pathway to lower embodied carbon via improved yields~\cite{eco-chip}, it must not come at the cost of system performance. Therefore, achieving true sustainability requires a holistic approach where carbon is treated as a primary optimization objective, co-optimized with traditional PPAC during the early stage architectural definition stages.

The complexity of HI systems requires co-design across the compute stack, including application, architecture (compute and memory), chip, and package space. 
This system-level co-design is highlighted in Fig.~\ref{fig:co-design-compute-stack}, which illustrates the variety of design considerations that must be made at each layer of the stack.
These considerations are also dependent on other decisions, blurring the boundaries between the layers of the stack. The entire design space is combinatorially large, making exhaustive search infeasible.

\begin{table*}[ht]
\centering
\caption{Comparison of prior modeling and optimization frameworks.}
\resizebox{\textwidth}{!}{%
\setlength{\arrayrulewidth}{1pt} 
\begin{tabular}{|l!{\vrule width 2pt}cccccccccc|}
\hline
\text{Framework}
& \shortstack{Chiplets of\\different sizes}
& \shortstack{Cycle-accurate\\performance} 
& \shortstack{D2D Protocol-\\aware} 
& \shortstack{Dataflow-\\aware} 
& \shortstack{Workload\\mapping} 
& \shortstack{Chiplet\\floorplanning} 
& \shortstack{DSE}
& \shortstack{Dollar cost}
& \shortstack{Embodied\\CFP} 
& \shortstack{Operational\\CFP} \\
\noalign{\hrule height 2pt}
\text{HISIM~\cite{hisim}} & \ding{55} & \ding{55} & \ding{55} & \ding{51} & \ding{55} & \ding{55} & \ding{51} & \ding{55} & \ding{55} & \ding{55} \\
\text{ACT~\cite{act-udit}} & \ding{55} & \ding{55} & \ding{55} & \ding{55} & \ding{55} & \ding{55} & \ding{55} & \ding{55} & \ding{51} & \ding{55}\\
\text{ECO-CHIP~\cite{eco-chip}} & \ding{51} & \ding{55} & \ding{55} & \ding{55} & \ding{55} & \ding{55} & \ding{55} & \ding{55} & \ding{51} & \ding{51} \\
\text{CATCH~\cite{puneet_catch_cost}} & \ding{51} & \ding{55} & \ding{55} & \ding{55} & \ding{55} & \ding{55} & \ding{55} & \ding{51} & \ding{55} & \ding{55} \\
\text{ChipletGym~\cite{chiplet-gym}} & \ding{55} & \ding{55} & \ding{55} & \ding{55} & \ding{55} & \ding{55} & \ding{51} & \ding{51} & \ding{55} & \ding{55} \\
\text{CarbonPATH~\cite{carbonpath-gh}}  & \ding{51} & \ding{51} & \ding{51} & \ding{51} & \ding{51} & \ding{51} & \ding{51} & \ding{51} & \ding{51} & \ding{51} \\
\hline
\end{tabular}%
}
\label{tab:prior_work}
\end{table*}

\noindent
{\bf Prior work.} A summary of prior work is provided in Table~\ref{tab:prior_work}. The table compares frameworks across key HI system co-design capabilities, spanning modeling assumptions, system awareness, and optimization scope. Specifically, we distinguish whether a framework supports  chiplets of different sizes  and whether it uses cycle-accurate performance models versus higher-level/analytical modeling. We also indicate whether the framework is aware of die-to-die (D2D) protocol constraints and dataflow characteristics. We highlight support for workload-to-chiplet mapping (exploring multiple mapping strategies) and chiplet floorplanning (optimizing chiplet placement and the implied D2D connectivity). We report whether each framework performs design-space exploration (DSE) and optimization, and whether it models economic and sustainability metrics, including dollar cost and embodied/operational carbon footprint (CFP). HISIM~\cite{hisim} presents an analytical modeling framework for the performance analysis of heterogeneous 2.5D/3D integrated AI systems, enabling efficient DSE. ~\cite{act-udit} leverages publicly available data to create data-driven models for embodied CFP, whereas ECO-CHIP~\cite{eco-chip} has modeled the overall CFP for chiplet-based systems, including both embodied and operational CFP (does only CFP analysis, no optimization). CATCH~\cite{puneet_catch_cost} provides cost models for multi-chiplet based designs, accounting for packaging, testing, and non-recurring engineering costs. ChipletGym~\cite{chiplet-gym} addresses the challenge of the vast design space by exploring it via the use of reinforcement learning (RL) and simulated annealing (SA). However, it falls short in several key aspects: (1) inaccurate modeling of PPAC due to unrealistic simplifying assumptions, (2) not including CFP as an optimization objective, (3) missing specific design considerations that lead to a narrower design space, (4) does not analyze the impact of diverse workload mapping styles, and (5) it fails to optimize across the entire stack. All the above aspects impact the design of HI systems.  

\noindent
{\bf CarbonPATH contributions.} 
CarbonPATH introduces a carbon-aware architecture exploration framework for chiplet-based AI systems that jointly considers workload mapping, chiplet architecture, and advanced packaging technologies. Unlike prior approaches that focus on performance or cost alone, CarbonPATH treats carbon footprint—both embodied and operational—as a first-class design objective and explicitly models its interaction with architectural and packaging decisions.

\begin{enumerate} 
    \item   \textbf{Carbon-aware cross-layer architecture exploration.}
We introduce the first framework that jointly explores workload mapping, chiplet architecture, and advanced packaging technologies while optimizing both embodied and operational CFP.

\item   \textbf{Topology-aware chiplet communication modeling.}
We develop analytical models that capture how chiplet geometry, bump pitch, packaging technology, and protocol overhead affect die-to-die bandwidth, latency, and energy.

    
\item 
\textbf{A scalable design-space exploration framework for sustainable AI hardware.}
CarbonPATH explores a large cross-layer design space spanning application mapping, architecture parameters, chiplet libraries, and packaging technologies, enabling system-level pathfinding for carbon-efficient AI accelerators.

\item 
\textbf{Design insights into carbon–performance tradeoffs for AI accelerators.}
Using CarbonPATH, we conduct a comprehensive exploration of chiplet-based AI accelerator designs across multiple workloads and packaging technologies. Our analysis reveals key tradeoffs between chiplet granularity, packaging style (2.5D vs. 3D), and workload characteristics, providing insights into how system designers can balance performance and lifecycle carbon footprint in future AI systems.



    

    
\end{enumerate}



Overall, CarbonPATH provides a methodology for suggesting architectures for GEMM workloads that optimize system-level PPAC and CFP. We open-source CarbonPATH at~\cite{carbonpath-gh}. 
\section{Preliminaries}
\label{sec:background}






\subsection{Advanced packaging architectures}
\begin{figure}[t]
\centering
\includegraphics[width=0.8\linewidth]{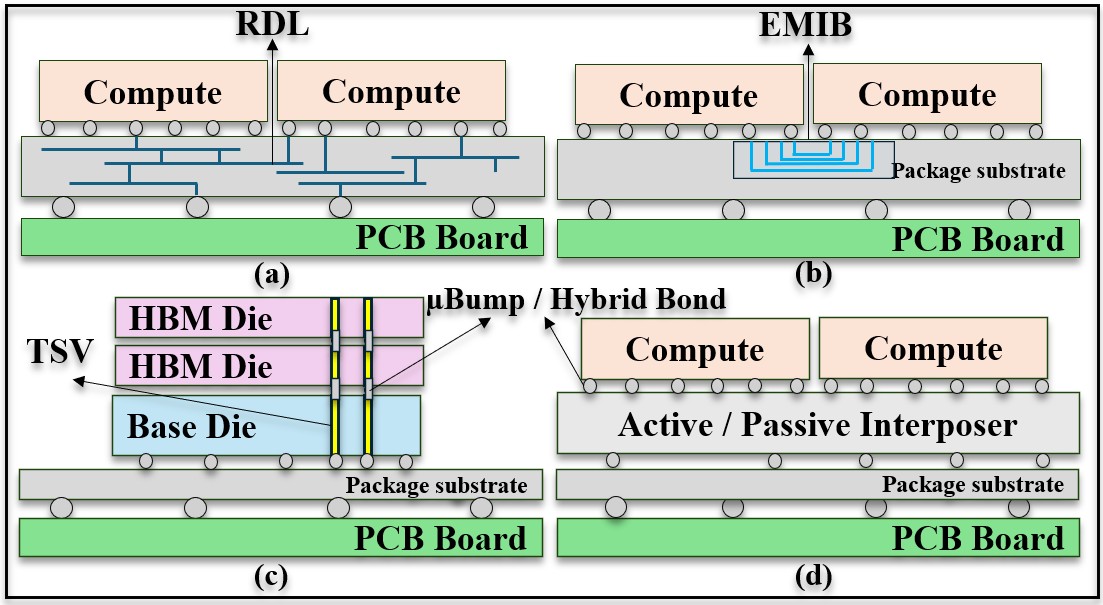}
\caption{Overview of different packaging interconnect architectures (a) RDL fanout, (b) EMIB, (c) TSV and \textmu Bump, and (d) hybrid bond.}
\label{fig:packaging-archs}
\end{figure}

It is increasingly complex for large monolithic SoCs to meet the performance demands of AI applications while managing cost as designs approach the reticle limit~\cite{chiplet-3, chiplet-4, chiplet-5, chiplet-2, chiplet-6}. To overcome these barriers, the industry is shifting toward a disaggregated design paradigm based on chiplets. In this approach, a monolithic system is partitioned into smaller modular dies that can be fabricated independently, often using the process technology best suited to their function, and then integrated within a single package. Three primary chiplet integration technologies (2.5D, 3D, and 2.5D+3D) are described below and illustrated in Fig.~\ref{fig:packaging-archs}.
Each technology has distinct tradeoffs in bandwidth, bump pitch, fabrication complexity, and cost.  Depending on the design goal, cost constraints, and data bandwidth requirements, there are multiple options~\cite{hi-1,hi-2,hi-3,hi-5} for 2.5D and 3D integration.

\subsubsection{2.5D integration technology}
2.5D integration places multiple chiplets side-by-side on a common substrate with fine-pitch die-to-die interconnects. Fig.~\ref{fig:packaging-archs} shows variants including redistribution layer (RDL) fan-out packaging (Fig.~\ref{fig:packaging-archs}(a)), which embeds chiplets in molding compound and connects them through 3–4 redistribution layers with 6/6 µm to 10/10 µm wiring~\cite{hi-1}; silicon bridge-based packaging such as Intel’s EMIB~\cite{emib-1} (Fig.~\ref{fig:packaging-archs}(b)), which embeds small silicon bridges with wiring down to 2 µm for localized high-bandwidth links~\cite{hi-1}; and silicon interposers (Fig.~\ref{fig:packaging-archs}(d)), which may be passive (BEOL only) or active (with FEOL support for routing or power delivery). These approaches provide scalable, high-bandwidth integration at lower cost than monolithic dies.

\subsubsection{3D integration technology}  
3D integration enables the vertical stacking of chiplets either directly on the package substrate, with connections formed using through-silicon vias (TSVs), microbumps, or hybrid bonding~\cite{tsv-1, hi-3, hi-2}. These technologies are highlighted in Fig.~\ref{fig:packaging-archs}(c). Face-to-back stacking relies on TSVs for tier-to-tier routing, while face-to-face stacking uses dense microbumps. Compared to 2.5D, 3D achieves shorter interconnects, higher bandwidth density, and lower energy per bit, but introduces greater complexity, yield challenges, and thermal issues due to vertical stacking.  

\subsubsection{2.5D+3D integration technology}  
2.5D+3D integration combines side-by-side and vertical packaging in a single system. Some chiplets are placed using 2.5D with interconnects such as RDL, EMIB, or silicon interposers, while others are stacked with 3D using TSVs, microbumps, or hybrid bonding. This approach enables tailoring of packaging style to workload needs—for example, stacking memory on compute dies for bandwidth, while placing logic side by side to meet thermal and cost constraints. By mixing 2.5D and 3D, these systems expand the design space and offer flexible trade-offs between performance, efficiency, and yield.

\subsection{Sustainability metrics}
\noindent
Design methodologies have now begun to explicitly quantify CFP (both operational and embodied), trading off performance gains against both manufacturing emissions and long-term operational costs. The works in~\cite{act-udit, eco-chip} have computed system CFP ($C_\text{sys}$), which consists of two terms and is given by~\cite{act-udit}:
\begin{equation}
\label{eq:system-carbon}
    C_\text{sys}= C_{\text{emb}} + C_{\text{ope}}
\end{equation}
\noindent
where $C_\text{emb}$ is the embodied CFP, and $C_\text{ope}$ is the operational CFP of the entire system under consideration.

\subsubsection{Embodied CFP}
\noindent
Embodied carbon refers to the emissions generated during the mining of materials, manufacturing, packaging, and transport. In the context of HI systems, this metric is driven by the complexity of the packaging technology and the specific technology nodes of the individual chiplets. Advanced nodes (e.g., 5nm, 3nm) require more energy-intensive lithography steps, increasing the carbon intensity per unit area~\cite{imec-wp}. Furthermore, yield plays a critical role; lower yields in complex bonding processes result in scrapped silicon, which amounts to wasted embodied carbon. Therefore, optimizing for embodied CFP requires balancing chiplet granularity, process node selection, and integration yield. The embodied CFP $C_\text{emb}$ of the entire system is given by~\cite{eco-chip}:

\begin{equation}
\label{eq:embodied}
    C_\text{emb}= \sum^{i=N_\text{c}}_{i=1} (C_{\text{mfg,i(n)}} + \frac{C_{\text{des,i}}}{N_\text{vol}}) + C_{\text{HI}}
\end{equation}

\noindent
where $N_\text{c}$ denotes the total number of chiplets, $C_\text{mfg,i(n)}$ is the CFP from manufacturing chiplet i in process node n, $C_\text{des,i}$ is the design stage CFP for chiplet i, amortized over production volume $N_\text{vol}$, and $C_\text{HI}$ is the CFP of advanced packaging incurred when assembling all $N_\text{c}$ chiplets at the package level. These metrics have been defined and modeled in~\cite{eco-chip}.



\subsubsection{Operational CFP}
\noindent
Operational carbon refers to the emissions generated during active use phase, determined by the system's power consumption and the carbon intensity of the energy grid. In the context of HI-based AI accelerators, this metric is a function of the energy dissipated by compute logic, memory access, and, crucially, the die-to-die (D2D) communication overhead inherent to heterogeneous systems. While decomposing a monolithic chip into chiplets improves manufacturing yield, the required data movement between dies introduces an energy penalty that increases the overall operational CFP. Because operational emissions accumulate over the system's entire operational lifetime, minimizing this metric requires a holistic approach: optimizing low-power architectures and efficient interconnect protocols. The operational CFP $C_\text{ope}$ for the entire system for a given workload is modeled as:
\begin{equation}
\label{eq:operational}
    C_\text{ope} = E_\text{system} * C_\text{src} * \text{Lifetime} * N_\text{vol} * T_\text{use}
\end{equation}
\noindent
where $E_\text{system}$ is the system energy that we model for HI systems as shown later in Eq.~\ref{eq:energy_model}, $C_\text{src}$ is the carbon intensity of the operating energy source, lifetime is the system lifetime that ranges from 3-7 years~\cite{lifetime-carbon, green-fpga, carbonset}, $N_\text{vol}$ is the production volume, and $T_\text{use}$ is the cumulative time the device is actively utilized over its lifetime. Detailed analytical models for this are available in~\cite{act-udit, eco-chip}.

\subsubsection{Performance sustainability index}
We utilize the performance sustainability index from~\cite{mobile-CFP}. The metric, inspired by performance per unit watt, helps analyze and optimize the tradeoff between performance and CFP and is defined as~\cite{mobile-CFP}:

\vspace{-3mm}
\begin{equation}
    \mathrm{Perf\text{-}SI} = \frac{\text{Performance}}{\text{CFP}}
    \label{eq:perfsi}
\end{equation}

\noindent 
where \textit{Performance} represents the system's latency for a given workload, and \textit{CFP} includes both the embodied and operational CFP.

\section{CarbonPATH framework overview}
\label{sec:overview}

\begin{figure}[t]
\centering
\includegraphics[width=\linewidth]{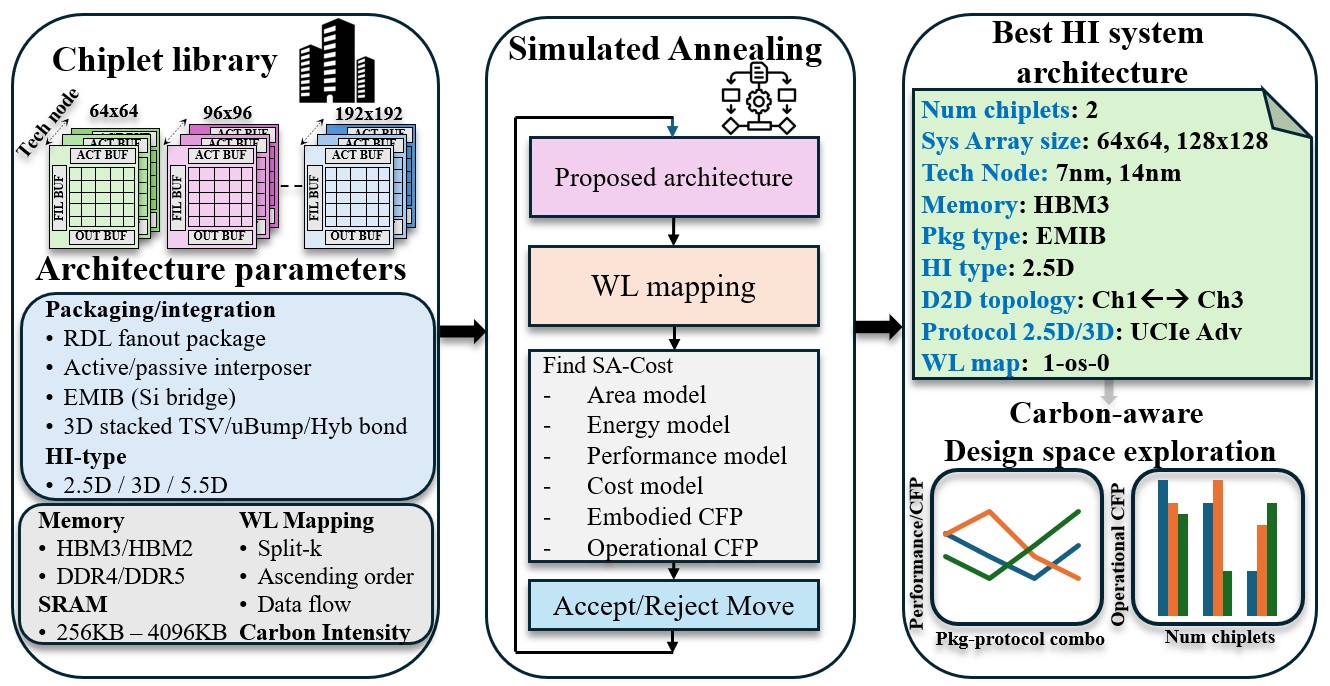}
\caption{CarbonPATH framework that leverages SA to identify the best HI system architecture for a given workload and design considerations.}
\label{fig:framework-path-ai}
\end{figure}

A top-level overview of the CarbonPATH framework is presented in Fig.~\ref{fig:framework-path-ai}. 
The goal is to identify optimized combinations of design choices across the application–architecture–chip–package space shown in Fig.~\ref{fig:co-design-compute-stack}. Given a GEMM workload and available design options, CarbonPATH uses SA to search the design space and select system parameters that minimize a cost function. 
We implement a library of systolic array–based chiplets across multiple array sizes, cache sizes, protocols, and technology nodes, each synthesized and characterized for area and power. Each chiplet is a pre-designed AI accelerator drawn from a chiplet library. CarbonPATH composes these chiplets into system-level models to estimate PPAC and CFP across multiple design configurations.
The SA cost function incorporates PPAC and CFP terms. Different optimization templates assign varying weights to these objectives in the cost function, enabling a systematic evaluation of tradeoffs. The figure illustrates the design space on the left, the SA engine in the center, and the optimized outputs on the right.


Table~\ref{tab:design-space} summarizes the parameter ranges explored in CarbonPATH. The left column lists the scope of the design space that CarbonPATH explores to identify optimized solutions, while the right column shows the specific parameter values used as inputs to our framework. These input values are configurable and can be adjusted to reflect different design scenarios. In the application and architecture space, parameters include workload mapping styles (\textit{assigning order, dataflow, and split-K}), the maximum number of chiplets in the system, and the choice of system memory (DDR or HBM variants). These determine how workloads are partitioned and scheduled across heterogeneous resources. The chip design space captures implementation-level choices, including technology nodes, systolic array sizes for compute chiplets, and SRAM buffer capacities matched to each array size. These parameters directly impact compute throughput, energy consumption, CFP, and silicon area. Finally, the packaging design space specifies integration styles (2D, 2.5D, 3D, and hybrid 2.5D+3D), interconnect types (RDL, EMIB, passive/active interposers, TSVs, microbumps, and hybrid bonding), and supported communication protocols (UCIe~\cite{ucie-pjb}, AIB~\cite{aib-pjb}, BoW~\cite{bow-pjb}). These determine the physical integration of chiplets and the bandwidth, latency, carbon, cost, and manufacturability of the HI system.


\begin{table}[t]
\centering
\caption{CarbonPATH design considerations grouped by application and architecture, chip, and packaging design spaces.}
\label{tab:design-space}
\resizebox{\linewidth}{!}{%
\setlength{\arrayrulewidth}{1pt} 
\begin{tabular}{|l|p{8cm}|}
\hline
\multicolumn{2}{|c|}{\textbf{Application \& chip architecture design space}} \\ \hline
Workload mapping & Assigning order, Dataflow (OS/WS/IS), split-K (on/off) \\ \hline
Maximum number of chiplets & 1--6 \\ \hline
System memory type & DDR4, DDR5, HBM2, HBM3 \\ \hline
\multicolumn{2}{|c|}{\textbf{Chiplet design space}} \\ \hline
Technology nodes & 7, 10, 14, 22, 28 \\ \hline
Systolic array sizes & 64$\times$64, 96$\times$96, 128$\times$128, 192$\times$192 \\ \hline


Chiplet: SRAM buffer size (KB) & \begin{tabular}[c]{@{}l@{}}64$\times$64:   256, 512, 768, 1024 \\ 
96$\times$96:   512, 1024, 1536, 2048 \\ 
128$\times$128: 1024, 2048, 3072, 4096 \\ 
192$\times$192: 2048, 4096, 6144, 8192 \end{tabular} \\ \hline

\multicolumn{2}{|c|}{\textbf{Packaging design space}} \\ \hline
Integration style & 2D, 2.5D, 3D, 2.5D+3D \\ \hline
Interconnect type & 2.5D: RDL, EMIB, Active interposer, Passive interposer \\ 
                   & 3D: TSV, Microbump, Hybrid bond \\ \hline
Protocols & 2.5D: UCIe\_STD, UCIe\_ADV, AIB, BoW \\
          & 3D: UCIe\_3D \\ \hline
Interconnect topology & Location of chiplet in 2.5D space and 3D space and its connections \\ \hline
\end{tabular}%
}
\end{table}

\section{CarbonPATH modeling framework}
\label{sec:modeling}
In CarbonPATH, we develop models to estimate PPAC for a given HI system configuration and utilize prior work for CFP estimation~\cite{eco-chip}.  CarbonPATH constructs system-level models and leverages ScaleSim~\cite{scalesim} to estimate latency and energy.

\subsection{Performance model}
\label{sec:model-latency}
To evaluate performance, we measure the latency of the computation. We assume the following data flow for the GEMM workload. The data is first read from DRAM into each chiplet—next, computation proceeds in parallel across chiplets. After computation, D2D communication transfers intermediate results to the destination chiplet, which we define as the largest chiplet since it typically offers the greatest compute capacity and memory bandwidth. The destination chiplet performs the reduction, and the result is written back to DRAM.

\noindent
\underline{System latency}
We model the latency for a specific workload, including compute latency, die-to-die (D2D) data transfer latency, and DRAM read and write access latency.  The overall latency for a specific workload on the HI system ($L_\text{system}$) is given by the sum of the maximum compute latency ($L_{\text{compute},i}$) and DRAM read latency ($L_{\text{DRAM\_RD},i}$) across all chiplets $N_c$, the D2D latency ($L_\text{D2D}$), and the maximum of write latencies ($L_{\text{DRAM\_WR},i}$) from the DRAM across all chiplets $N_c$. Since workloads execute in parallel across chiplets, we take maximum values for compute and DRAM operations. Thus, $L_\text{system}$ is given by:

\begin{align}
L_{\text{system}} &= 
\underbrace{\max_{1 \le i \le N_c} \big( L_{\text{compute},i} + L_{\text{DRAM\_RD},i} \big)}_{\text{Compute + DRAM Read}} \nonumber\\
&\quad + \underbrace{L_{\text{D2D}}}_{\text{D2D}} + \underbrace{\max_{1 \le i \le N_c} \big( L_{\text{DRAM\_WR},i} \big)}_{\text{DRAM Write}}
\end{align}

\noindent
\underline{Compute latency and workload mapping.} 
We obtain compute latency using \textit{ScaleSim} simulations.
In the simulation, each systolic array is parameterized with a specified main memory bandwidth, three equally sized on-chip memory buffers, an assigned data flow type, and a list of scheduled GEMM tiles. We calculate the latency using the frequency (obtained after logic synthesis in a 7nm technology node) and simulated cycle count.

The system scheduler is responsible for the workload mapping and scheduling, which also impacts latency. The detailed workload scheduling algorithm is presented in Algorithm~\ref{alg:mapping_algorithm}. Scheduler first receives three mapping parameters, including \textit{split-K}, \textit{assigning order}, and \textit{dataflow}. Then it partitions the overall workload into small tiles and assigns them to the systolic arrays according to their relative compute throughput. \textit{split-K} determines if the K-dimension (the reduction dimension in a GEMM workload) is partitioned. When enabled, it introduces significant interconnect traffic due to the aggregation of partial sums, which are reduced on the largest core and written back to main memory. \textit{Assigning order} determines whether workloads are allocated to the smallest core or the largest core first. This order may impact the fairness of workload distribution, particularly in heterogeneous architectures. \textit{Dataflow} is passed to \textit{ScaleSim} and is one of output stationary (OS), input stationary (IS) or weight stationary (WS)~\cite{scalesim}. The~\textit{dataflow} type plays a critical role in determining overall performance, particularly in conjunction with the shape of the workload.

\begin{algorithm}[t]
\caption{Workload tiling and assignment.}
\label{alg:mapping_algorithm}
\begin{algorithmic}[1]
\Require Workload dimensions $(M, K, N)$; tile sizes $(t_M, t_K, t_N)$; flags $s_K$, $s_A$; cores $\{c_p\}_{p=1}^P$ with compute powers $\{p_p\}_{p=1}^P$; dataflow $d \in \{\mathrm{OS}, \mathrm{WS}, \mathrm{IS}\}$
\Ensure Tile assignment mapping $\mathcal{A}(c_p)$ for each core
\State Base tile sizes: $b_M \gets t_M$, $b_N \gets t_N$, $b_K \gets t_K$ if $s_K$ else $K$
\State Sort cores by $p_p$ in ascending order if $s_A$, otherwise descending; let sorted order be $(c_{(1)}, \ldots, c_{(P)})$
\State Partition $M$ into $\{m_i\}_{i=1}^{I}$, $K$ into $\{k_j\}_{j=1}^{J}$, and $N$ into $\{n_\ell\}_{\ell=1}^{L}$, using respective base sizes; allow last tiles to exceed base size if necessary
\State Construct tile set $\mathcal{T} = \{(m_i, k_j, n_\ell)\} \quad \forall i \in [1{:}I],\; j \in [1{:}J],\; \ell \in [1{:}L]$; total number of tiles $T = I \cdot J \cdot L$

\For{$p = 1$ to $P$}
    \State Compute $r_p \gets p_p / \sum_{q=1}^{P} p_q$;\quad ideal tile count $d_p \gets r_p \cdot T$
    \State Assign $\hat{n}_p \gets \lfloor d_p \rfloor$
\EndFor
\State Distribute remaining $R = T - \sum_{p=1}^{P} \hat{n}_p$ tiles to cores with the largest fractional parts $(d_p - \lfloor d_p \rfloor)$
\State Initialize index $S \gets 1$
\For{$p = 1$ to $P$}
    \State Assign dataflow $d$ and map tiles: $\mathcal{A}(c_{(p)}) \gets \mathcal{T}[S : S + \hat{n}_{(p)} - 1]$
    \State Update $S \gets S + \hat{n}_{(p)}$
\EndFor
\end{algorithmic}
\end{algorithm}

\begin{figure}[t]
\centering
\includegraphics[width=0.5\linewidth]{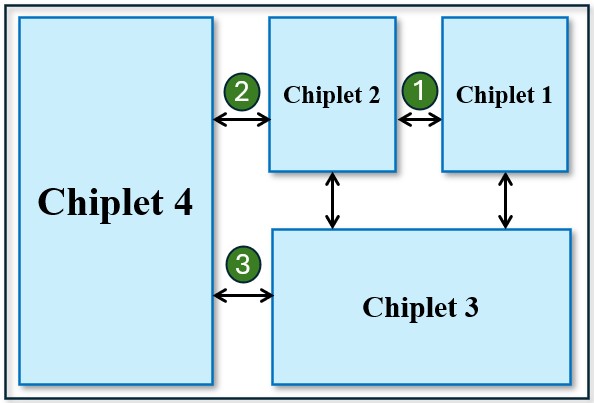}
\caption{Example to demonstrate the topology and datapath for latency estimation in 2.5D integration technology.}
\vspace{-8mm}
\label{fig:d2d_show}
\end{figure}

\noindent
\underline{Die-to-die (D2D) latency} CarbonPATH models two types of D2D chiplet communication: compute–compute and compute–memory. 

\noindent
\textit{(1) Communication between compute chiplets: } 
CarbonPATH models D2D latency between compute chiplets by accounting for congestion, traffic paths, and topology that reflect realistic interconnect constraints.  For both 2.5D and 3D integration, the D2D latency between compute chiplets is modeled as  $L_{\text{d2d}, i \rightarrow j} = \frac{\text{Data}}{BW_{\text{d2d}, i \rightarrow j}}$
where the effective bandwidth is $ BW_{\text{d2d}, i \rightarrow j} = \min(BW_{\text{d2d}, i}, BW_{\text{d2d}, j})$. Here, $\text{Data}$ is the transfer size (bits) and  $BW_{\text{d2d}, i \rightarrow j}$ is the effective bandwidth between two adjacent dies $i$ and $j$,  which depends on packaging technology, interconnect topology, protocol overheads, and the number of I/Os. The $BW_{\text{d2d}, i}$ is the maximum bandwidth supported by chiplet $i$.

This ensures that the weakest link dominates the path, while the topology captures congestion and shared interconnect usage. We explain the topology of the interconnect using Fig.~\ref{fig:d2d_show} as an example. The figure illustrates a four-chiplet system in which adjacent chiplets are interconnected. Based on floorplanning results from our area model (Sec~\ref{sec:model-area}), we identify neighboring chiplets. The largest chiplet is designated as the \textit{destination} to aggregate partial results (e.g., in GEMM reduction). In the example, Chiplet 4 serves as the destination. Data from Chiplets 1–3 is routed to Chiplet 4 through adjacent links, with sequential transfers assumed when common links are shared. The number of memory channels is determined by the size of the compute chiplet it interfaces with, since larger chiplets can accommodate more IOs for memory read and write operations.  
We estimate the maximum die-to-die bandwidth for a chiplet as follows:




\begin{equation}
    BW_{\text{d2d}, n} = \text{DR}_{n} \times N_{\text{bump}} \times \eta_{\text{protocol}}
    \label{eq:d2d_bw_model1}
\end{equation}

Here, $BW_{\text{d2d}, n}$ is the maximum die-to-die bandwidth for a given chiplet $n$, $N_{\text{bump}}$ is the total number of bumps in the chiplet that can be used for D2D communication, dictated by package characteristics such as bump pitch and chip area. The maximum data rate per bump, $\text{DR}_{n}$ is constrained by the communication protocol and its physical layer (PHY). Different protocols are compatible with different packaging technologies, as summarized in Table~\ref{tab:mapping_pkg_protocol}.   The factor $\eta_{\text{protocol}}$ captures protocol efficiency, i.e., the fraction of the raw physical data rate that is usable for payload transfer after accounting for encoding, framing, and link management overheads. The number of bumps $N_\text{bump}$ is given by:

\begin{equation}
N_{\text{bump}} = 
\begin{cases}
\left\lfloor \dfrac{A_{\text{chiplet}}}{B_{\text{pitch}}^2} \right\rfloor, & \text{3D integration} \\[1em]
\left\lfloor \dfrac{L_{\text{chiplet}}}{B_{\text{pitch}}} \right\rfloor,   & \text{2.5D integration}
\end{cases}
\end{equation}

\noindent
Here, $A_{\text{chiplet}}$ is the chiplet area, $L_{\text{chiplet}}$ is the chiplet perimeter length available for IOs, and $B_\text{pitch}$ is the bump pitch.   In 3D integration, bumps can be distributed across the full die area, whereas in 2.5D integration the bumps available for die-to-die communication are usually constrained to the chiplet edges. For 2.5D+3D hybrid integration, some chiplets are interconnected using edge-limited bumps (2.5D model) while others employ area-limited bumps (3D model). This peripheral placement in 2.5D integration is necessary by the strict requirements of D2D communication IPs, which demand tight length matching and minimal routing congestion. Constraining bumps to the edge avoids routing signals through the central die area, which is dominated by dense Vdd/Vss power grids and clock meshes, thereby preventing cross-core logic crossings that would otherwise degrade power and signal integrity.

\noindent
\textit{(2) Communication between compute and memory chiplets:} CarbonPATH models compute-to-memory communication differently for 2.5D and 3D integration.  For 2.5D integration, memory bandwidth is distributed across chiplets, with larger chiplets assigned more channels and thus higher bandwidth. Since each chiplet can access DRAM independently, DRAM reads and computations proceed in parallel. The DRAM read latency for chiplet $i$ is modeled as $  L_{\text{DRAM\_RD}, i} = \frac{\text{Data}}{BW_{\text{mem}, i}}$  where $BW_{\text{mem}, i}$ is fixed based on the chiplet size. 

For 3D integration, only the base die has direct access to DRAM, while other dies communicate with the DRAM via the base die. As a result, D2D links constrain the effective bandwidth of stacked dies. For the base die, latency follows the same model as 2.5D. For other dies, the effective bandwidth is the minimum of the DRAM bandwidth and all D2D bandwidths along the access path. 
For example:  
\begin{align}
    BW_{\text{eff, Die 1}} &= BW_{\text{dram}}, \\
    BW_{\text{eff, Die 2}} &= \min(BW_{\text{dram}}, BW_{\text{d2d}, 1 \rightarrow 2}), \\
    BW_{\text{eff, Die 3}} &= \min(BW_{\text{dram}}, BW_{\text{d2d}, 1 \rightarrow 2}, BW_{\text{d2d}, 2 \rightarrow 3}), \\
    & \ \ \vdots
\end{align}

The write latency depends on the \textit{split-K} parameter, which determines whether all chiplets write back results individually or only the destination chiplet performs the final write. If \textit{split-K} is disabled, each chiplet writes back its own computation result to the DRAM independently so  $L_{\text{DRAM\_WR}} = \max_{i \in \mathcal{C}}  \left(\frac{\text{Data}_{\text{wr},i}}{BW_{\text{mem},i}}\right)$.  If \textit{split-K} is enabled, only the destination chiplet performs the write after reduction. $
        L_{\text{DRAM\_WR}} = \frac{\text{Data}_{\text{WR},d}}{BW_{\text{mem},d}}$
where $d$ is the destination chiplet.   For memory chiplets the maximum bandwidth is derived from standard memory specifications such as DDR4, DDR5, HBM2, and HBM3~\cite{ddr-hbm-spec}

\begin{table}[t]
\centering
\caption{Compatible combinations of integration and packaging interconnect type, and communication protocols.}
\footnotesize 
\begin{tabular}{!{\vrule width 2pt}c|c!{\vrule width 2pt}c|c!{\vrule width 2pt}}
\noalign{\hrule height 2pt} 
\textbf{Package} & \textbf{Protocol} & \textbf{Package} & \textbf{Protocol} \\ 
\noalign{\hrule height 2pt} 

2.5D RDL & UCIe-S  & 3D TSV & UCIe-3D    \\ \hline
2.5D EMIB & UCIe-A, AIB, BoW  &   3D uBump & UCIe-3D    \\ \hline
2.5D Passive & UCIe-A, AIB, BoW &   3D HybBond & UCIe-3D    \\ \hline 
2.5D Active & UCIe-A, AIB, BoW  & &   \\ \hline

\noalign{\hrule height 2pt} 
\end{tabular}
\label{tab:mapping_pkg_protocol}
\end{table}

\subsection{Energy or power model}
\label{sec:model-energy}
The system energy is modeled as the sum of compute energy, and die-to-die (D2D) communication energy (includes compute-compute and compute-memory)
\begin{equation}
    E_{\text{system}} = E_{\text{compute}} + E_{\text{d2d}} 
    \label{eq:energy_model}
\end{equation}

\noindent
\underline{Compute and memory access energy.}  
This energy accounts for compute operations, on-chip SRAM activity, and DRAM accesses across all chiplets and is given by:  
\begin{equation}
    E_{\text{compute}} = \sum_{i=1}^{N_C} 
    \big(E_{\text{DRAM\_RD},i} + E_{\text{compute},i} + E_{\text{SRAM},i}\big) 
    + \sum_{i=1}^{N_C} E_{\text{DRAM\_WR},i}
    \label{eq:energy_compute}
\end{equation}


Here, $N_\text{c}$ is the number of chiplets, $E_\text{DRAM\_RD,i}$ and $E_\text{DRAM\_WR,i}$ denotes the energy consumed by chiplet $i$ for off-chip DRAM reads and writes, respectively (i.e., data movement between DRAM and the chiplet). $E_\text{SRAM,i}$ is the energy due to on-chip SRAM access (reads/writes to local buffers) within chiplet $i$. $E_\text{compute,i}$ captures the energy of compute operations performed on each chiplet $i$, excluding the memory-access energy. Each term in Eq.~\ref{eq:energy_compute} is modeled by $E_{\mathrm{x}}^{(bit)} \times B_{(x)}$, where $E_{\mathrm{x}}^{(bit)}$ is the energy consumed per bit (pJ/bit) for activity $x \in \{\mathrm{DRAM\_RD}, \mathrm{DRAM\_WR}, \mathrm{SRAM}, \mathrm{compute}\}$, obtained from prior characterization~\cite{seo-sram-area-power,dram-energy-1,dram-energy-2,bow-pjb,ucie-pjb,aib-pjb}, and $B_{(x)}$ is the total number of bits transferred/accessed/processed for that activity as determined by the simulator at runtime. For DRAM writes, the formulation naturally handles both cases of the \textit{split-K} parameter. If \textit{split-K} = 1, only the destination chiplet ($i$) performs the final DRAM write, so $E_{\text{DRAM\_WR},i} > 0$ while all others are zero. If \textit{split-K} = 0, all chiplets independently write to DRAM, and each term $E_{\text{DRAM\_WR},i}$ is non-zero. DRAM read and write energy per bit are obtained from~\cite{dram-energy-1,dram-energy-2}, and SRAM read and write energy per bit are obtained from~\cite{seo-sram-area-power}.


\noindent
\underline{D2D communication energy.}  
D2D energy is computed from the total number of bits transferred across all die-to-die links:  
\begin{equation}
    E_{\text{d2d}} = \sum_{(i,j) \in \mathcal{L}} E_{\text{bit,d2d}}^{i \rightarrow j} 
    \times \text{total\_bits}_{i \rightarrow j}
    \label{eq:energy_d2d}
\end{equation}

where $\mathcal{L}$ is the set of D2D links, $E_{\text{bit,d2d}}^{i \rightarrow j}$ is the per-bit energy cost of the interconnect between chiplets $i$ and $j$, and $\text{total\_bits}_{i \rightarrow j}$ is the amount of data exchanged which is calculated by model described in Sec~\ref{sec:model-latency} (Die-to-die (D2D) latency). The D2D link energy per bit is sourced from~\cite{bow-pjb,ucie-pjb,aib-pjb}.

\subsection{Area model}
\label{sec:model-area}
To model the area footprint ($A_\text{system}$), defined as the total two-dimensional space occupied by the system, we first identify the type of system under evaluation. For a monolithic system, the area footprint is simply the area of the design itself from the chiplet library, as no additional integration area is required. For a 3D system, the area footprint is defined by the area of the base die. For 2.5D and 2.5D+3D systems, the package or interposer area is estimated using a simple bipartitioning-based slicing chiplet floorplanner~\cite{eco-chip,recursive-book}. The algorithm hierarchically organizes the chiplets within a bounding box by recursively partitioning the set of chiplets and making alternate vertical and horizontal cuts. It creates bi-partitions that are closely balanced. The recursive algorithm alternates its slicing direction (i.e., vertical and horizontal cuts) and assumes a rectangular aspect ratio. The recursion terminates when only a single chiplet remains in a partition. The 2.5D and 2.5D+3D systems incur white-space or area overhead, which is computed after floor planning.

\subsection{Dollar cost model}
\label{sec:model-cost}
The monetary cost of an HI system is modeled as follows~\cite{puneet-chiplet-small-cost, chiplet-costonly}

\begin{equation}
M_{\text{system}} = 
\frac{\sum_{i=1}^{N_C} M_{\text{chiplet}_i} + M_{\text{interposer}} + M_{\text{pkg}}}
{Y_{\text{bonding}}} + M_{\text{mem}}
\end{equation}

\noindent
where $M_{\text{chiplet}_i}$ is the cost of the individual $\text{chiplet}_i$, $N_C$ is the number of chiplets in the system, $M_{\text{interposer}}$ is the cost of the interposer, $M_{\text{pkg}}$ is the cost of the package substrate, $M_{\text{mem}}$ is the cost of the memory, and $Y_{\text{bonding}}$ is the bonding yield that depends on the integration and packaging interconnect in the HI system~\cite{3d-carbon}. Interposer and substrate costs are derived from floorplanned area. For 2.5D RDL or silicon-bridge interconnects, $M_{\text{interposer}} = 0$, while for 2.5D active or passive interposers, it is modeled as the cost of an additional silicon chiplet fabricated in an older technology node (65nm)~\cite{eco-chip,3d-carbon}. Package substrate cost is area-dependent based on~\cite{chiplet-costonly}, and memory cost is from~\cite{mem_cost}. 
The chiplet cost $M_{\text{chiplet}}$ is determined as follows~\cite{puneet-chiplet-small-cost}:

\begin{equation}
M_{\text{chiplet}} =  
\frac{M_{\text{wafer}, \, \text{chiplet}}}{DPW_{\text{chiplet}_i}} 
\cdot \frac{1}{Y_{\text{chiplet}_i}}
\end{equation}

\noindent
where $M_{\text{wafer}, \, \text{chiplet}}$ is the cost of the wafer for manufacturing the chiplet, $DPW_{\text{chiplet}}$ is the number of dies per wafer for the chiplet, and is modeled as in~\cite{eco-chip}, and $Y_{\text{chiplet}}$ is the yield for the chiplet modeled  using the negative binomial distribution~\cite{yield-1,yield-2, yield-3}:





\subsection{CFP model}
\label{sec:model-carbon}

\noindent
The CFP models are adopted from~\cite{eco-chip} as described in Sec~\ref{sec:background}. We estimate the total CFP using Eq.~\ref{eq:embodied} and~\ref{eq:operational} for estimating the embodied and operational CFP.

\section{CarbonPATH Optimization Framework}
CarbonPATH employs a standard simulated annealing (SA) approach to identify the optimized HI system configuration that minimizes the overall SA-defined cost for a given workload. As with any SA framework, we define three core components: (1) solution space, (2) moves, and (3) cost function. We also discuss optimizations performed for improving the efficiency and runtime of the SA engine. 

\subsection{Solution space}

The solution space for our SA engine is defined as the comprehensive set of all possible architectural configurations derived from the combinatorial interplay between application requirements, chip architecture, package-level technologies, and chiplet-specific specifications, as detailed in Table~\ref{tab:design-space}. In this context, design space enumeration involves systematically identifying all potential combinations of these multi-level variables, where each unique design point—or candidate solution—is formally represented as a high-dimensional vector of design parameters. Within this vector, each element represents a specific design decision, such as total chiplet count, technology node, die-to-die interconnect protocol, etc. Combining these parameters from all levels leads to exponential growth in the design space, creating a vast number of potential configurations to explore. 

Table~\ref{tab:mapping_pkg_protocol} outlines the supported package-protocol configurations. In the 2.5D space, RDL supports the UCIe-S protocol, while EMIB, Passive, and Active substrates each support UCIe-A, AIB, and BoW, resulting in 10 distinct combinations. For 3D integration, TSV, $\mu$Bump, and hybrid bonding each pair with UCIe-3D to provide 3 options. Finally, CarbonPATH accounts for heterogeneous 2.5D+3D systems by combining every valid 2.5D configuration with each 3D option, an additional 30 combinations are evaluated. To enable a comprehensive design space exploration, CarbonPATH evaluates 43 diverse interconnect and protocol pairs alongside 12 workload mapping strategies. These 12 workload mapping strategies are further detailed in Table~\ref{tab:design-space}. The goal is to efficiently explore this space while only permitting feasible combinations of design parameters. Our SA-based framework, therefore, searches exclusively through valid design choices. 
To ensure the exploration of valid design spaces, invalid architectural configurations are strictly prohibited. This includes mismatched protocol assignments (e.g., UCIe-3D in a 2.5D system), unstable 3D stacks (e.g., stacking a larger die onto a smaller one), or incorrect classifications (e.g., assigning a 2.5D+3D HI type to a simple system with two chiplets). Our framework enforces these constraints through validation checks at initialization and following every architectural transformation, ensuring that all HI-system parameter vectors remain feasible.
At the start of simulated annealing, CarbonPATH generates a random but \emph{valid} HI system architecture drawn from the feasible design space. All subsequent perturbations then modify parameters at one of the following levels: \textit{application}, \textit{chip architecture}, \textit{package}, or \textit{chiplet}. 

\subsection{Simulated annealing moves}
The exploration of the design space and optimization occurs through \emph{moves}. At each iteration, the SA engine perturbs the current configuration to generate a new candidate solution, evaluates its cost, and then decides whether to accept or reject the move based on the cost difference and the annealing schedule. The set of moves must be defined such that the entire solution space remains searchable.

We adopt a hierarchical move selection strategy in which CarbonPATH first chooses whether to apply an application-level perturbation (workload mapping) or a lower-level perturbation (architecture, chiplet, or package). This hierarchical approach reflects the multi-scale nature of the design space.

\noindent
\underline{Application-level moves:} At the application level, a move perturbs how the workload is mapped onto the HI system. We have three possible moves at this level. The first is modifying \textit{dataflow} in the systolic array (e.g., switching between output-stationary (OS), weight-stationary (WS), or input-stationary (IS) modes). The second changes how the workload is partitioned across chiplets, such as by splitting along the $K$-dimension (\textit{split-K}). 
The third move is \textit{assigning order}, which specifies the assignment order that determines how sub-workloads are distributed across chiplets.

\noindent
\underline{Chip-architecture-level moves:} At the chip-architecture level, we have two possible moves. The first move either increases or decreases the total number of chiplets in the system relative to the previous iteration. When chiplets are added or removed, compliance checks and corrective modifications ensure the design remains consistent with architectural and protocol rules (e.g., a 3D stack requires at least two chiplets). When a change in chiplet count renders the current HI-package integration type invalid, the framework dynamically adjusts the HI-package configuration among 2D, 2.5D, 3D, and 2.5D + 3D types based on the total chiplet count. This ensures that the package integration remains consistent with the updated system while strictly adhering to architectural feasibility rules. The second move changes the type of memory in the architecture (DDR vs. HBM) to a different option from Table~\ref{tab:design-space}, thereby altering performance and cost trade-offs.  

\noindent
\underline{Chiplet-level moves:} At this level, we perform one move that replaces a selected chiplet with another from the chiplet library. The library provides variants across technology nodes, systolic array sizes, and SRAM buffer sizes. This move inherently perturbs one or more of these parameters. The replacement is accepted only if validity constraints (e.g., no infeasible stacking such as placing a larger die beneath a smaller one) are satisfied.

\noindent
\underline{Package-level moves:} At the package level, we have two possible moves that perturb packaging and interconnect options. The first move changes the package interconnect type while maintaining the same HI integration style (e.g., 2.5D vs. 3D). The second move modifies the D2D communication protocol, selecting an alternative supported by the current HI system from the chiplet library. These moves enable the exploration of the integration and communication trade-offs inherent in advanced packaging technologies.

\subsection{Cost function}
A central component of the SA framework is the cost function, which guides the search through the design space by quantifying the quality of each candidate system configuration. The cost function integrates multiple design objectives into a single scalar metric, enabling the SA engine to compare alternative solutions and make accept/reject decisions during the optimization process. In CarbonPATH, the cost function captures energy (power), area, latency (performance), embodied CFP, operational CFP and dollar cost considerations, as defined below.
\vspace{-3mm}
\begin{equation}
    \text{SA-Cost} = \alpha E_\text{system} + \beta A_\text{system} + \gamma L_\text{system} + \theta M_\text{system} + \zeta C_\text{emb} + \eta C_\text{ope}
    \label{eq:sa_cost}
\end{equation}

The SA cost function combines key metrics (latency, energy, area, dollar cost, embodied, and operational CFP) defined by the models in Sec~\ref{sec:modeling}. 
Here $\alpha$ is energy coefficient, $\beta$ is area coefficient, $\gamma$ is latency coefficient, $\theta$ is the cost coefficient, $\zeta$ is embodied CFP coefficient, and $\eta$ is the operational CFP coefficient. To enable a comprehensive exploration of the architectural design space, it is essential to incorporate both embodied and operational CFP into the cost function. We add $\zeta$, and $\eta$ coefficients for the CFP components to facilitate flexible user prioritization, the analysis of which is detailed in Sec~\ref{sec:results-template-optimizations}. As shown in Eq.~\ref{eq:embodied} and~\ref{eq:operational}, the relationship between physical metrics (area, energy) and CFP is non-linear, i.e., embodied CFP depends on area and yield, and yield has an exponential dependence on area. Furthermore, operational CFP is dynamic, governed by location-specific carbon intensity, device lifetime, and deployment volume. 

Since the metrics have different units and scales, normalization is necessary to prevent any single term from dominating the cost function. To achieve this, CarbonPATH evaluates 10,000 randomly generated valid HI system architectures to obtain the distribution of each metric. For each term, we normalize by subtracting the minimum observed value and dividing by the observed distribution's median. This ensures that all metrics contribute comparably to the overall cost while preserving relative trade-offs.   The weighting factors $\alpha, \beta, \gamma, \theta, \zeta$ and $\eta$ allow CarbonPATH to balance these objectives according to user-defined design priorities.


\subsection{Runtime considerations}
\label{sec:sa-runtime-considerations}
SA is typically runtime-intensive, with execution time dominated by repeated evaluations of the cost function. In CarbonPATH, the cost function is largely based on analytical models (Sec~\ref{sec:modeling}), which are fast to evaluate. However, latency and energy metrics rely on cycle-accurate simulations from \textit{ScaleSim}, which are significantly more computationally expensive. To mitigate this bottleneck, we implement two runtime optimizations.

First, we adopt an incremental cost computation approach. Certain moves do not alter the cycle count of the systolic array and therefore do not require a full re-simulation. For example, changing the technology node of a chiplet only affects the achievable frequency and energy, but leaves the number of cycles unchanged. In such cases, latency can be updated analytically from the previous simulation result without rerunning \textit{ScaleSim}. By contrast, moves such as workload mapping perturbations directly affect communication and data reuse patterns, thereby altering the number of cycles and requiring a re-simulation. Second, we introduce a lookup-table-based \textit{simulation cache}. Each execution of \textit{ScaleSim} records key parameters of the simulated systolic array, including workload shape, main memory bandwidth, on-chip buffer size, dataflow, and cycle count. During the SA process, a full simulation is only triggered if the parameter configuration has not been previously encountered; otherwise, cached results are retrieved. This reduces redundant simulations and provides a substantial runtime speedup. 

\section{CarbonPATH Analysis}
\label{sec:results}
\noindent

\subsection{Experimental setup and methodology}

\noindent
The CarbonPATH framework runs by taking a GEMM workload as input to the SA optimization framework.  We evaluate the CarbonPATH framework on multiple representative GEMM workloads from modern DNNs, as shown in Table~\ref{tab:gemm_dims}. For each workload, we consider multiple optimization templates shown in Table~\ref{tab:optimization_profile_table}.

\noindent
\underline{Parameter values:} CarbonPATH relies on multiple input parameters based on the model. Chiplet area and power are obtained by synthesizing RTL for various systolic array sizes using Synopsys Design Compiler with  ASAP7 PDK~\cite{ASAP7}. Values for other technology nodes are scaled according to~\cite {eco-chip}, assuming a 12.5\% activity factor for power analysis. SRAM energy for different memory capacities is taken from~\cite{seo-sram-area-power}. Latency is evaluated using ScaleSim \cite{scalesim} at 1 GHz for 7nm, consistent with synthesis, and scaled for other technology nodes as in~\cite{freq-scaling}. We evaluate the embodied and operational CFP from ~\cite{eco-chip}, modifying the framework to support new package and system types. For die-to-die energy per bit, we adopt the power efficiency values from~\cite{bow-pjb,ucie-pjb,aib-pjb}, and for DRAM energy, we use~\cite{dram-energy-1,dram-energy-2}. Dollar cost for compute and memory chiplet manufacturing and integration is modeled using wafer cost estimates from~\cite{cost-per-wafer,mem_cost}. In our analysis, we assume a production volume of 1 million; all of these parameters are configurable knobs that users can modify as needed.   The SA hyperparameters used are an initial temperature of 4000, a final temperature of 0.001 with a cooling rate of 0.99, and 50 moves per temperature. 

\noindent
\underline{Notation:} 
For chiplet attributes, we use the format \textit{A-T-S}, where \textit{A} specifies array size, \textit{T} specifies technology node, and \textit{S} specifies memory capacity/size. E.g., a chiplet with a 64$\times$64 systolic array implemented in 7nm and equipped with a 512 KB SRAM buffer is denoted as \textit{64-7-512}.

\noindent
For workload mapping strategies, we use the format \textit{O-D-K}, where \textit{O} denotes the ~\textit{assigning order}, \textit{D} specifies the \textit{dataflow}, and \textit{K} indicates whether \textit{split-K} is enabled. For example, \textit{1-OS-0} represents ~\textit{assigning order} set to 1, output-stationary \textit{dataflow}, and \textit{split-K} disabled.   

\noindent
For packaging, we use the format \textit{I-P-M}, where \textit{I} specifies the integration style (2.5D, 3D, or 2.5D+3D), \textit{P} denotes the packaging interconnect (RDL, EMIB, TSV, active interposer, passive interposer, \textmu bump, or hybrid bond), and \textit{M} indicates the memory type (DDR, HBM). For instance, \textit{2.5D-RDL-DDR5} refers to a 2.5D integration using RDL fan-out wafer-level packaging with DDR5 memory. For brevity, we also use compact abbreviations: \textit{Acti} and \textit{Pass} for active and passive interposers, and \textit{\textmu B} and \textit{HB} for \textmu bump and hybrid bond.  

\noindent
For protocols, we adopt short notations such as UCIe 3D \textit{(UC3)}, UCIe Standard \textit{(UCS)}, and UCIe Advanced \textit{(UCA)}.



We run CarbonPATH across multiple workloads covering the full range of optimization templates shown in Table~\ref{tab:optimization_profile_table}. We then compare these results with ChipletGym~\cite{chiplet-gym}. These cost functions are input to our framework, and users have the control to modify them to optimize the required parameter. For the rest of the paper, we use specific terminology to describe our system configuration in our experiments and results. The term \textit{identical chiplet system} refers to an HI system comprising four identical chiplets, each in a~\textit{128-7-1024} configuration, and the term \textit{different chiplet system} denotes an HI system utilizing four distinct chiplets with configurations of~\textit{64-7-256, 96-7-512, 128-7-1024,} and ~\textit{192-7-2048}. In both cases, unless specified, we use a DDR5 DRAM memory architecture and a workload mapping of~\textit{1-OS-0} for workload 1\footnote{We observe similar results across other workloads and configurations and present these as representative examples}.

The remainder of this section presents an analysis of traditional PPAC metrics in Sec~\ref{sec:result-ppac}. We then discuss the carbon implications of the HI system in Sec~\ref{sec:carbon-results}. Followed by Sec~\ref{sec:result-sa-overall} with SA optimization results, comparisons to existing works, and results with various templates.



\begin{table}[t]
\centering
\caption{GEMM workloads for different benchmarks.}
\resizebox{\linewidth}{!}{%
\begin{tabular}{!{\vrule width 2pt}c|l!{\vrule width 2pt}c|c|c!{\vrule width 2pt}}
\noalign{\hrule height 2pt} 
\textbf{WL} & \textbf{Benchmark - Layer} & \textbf{M (Batch)} & \textbf{K (Input)} & \textbf{N (Output)} \\
\noalign{\hrule height 2pt} 
1 & GPT-2 - MLP (feed-forward)  & 512 & 768  & 3072  \\ 
2 & ViT - MLP (batch=32)      & 6304  & 768  & 3072  \\ 
3 & ViT - MLP (batch=1)       & 197  & 768 & 3072  \\ 
4 & ResNet-50 - FC (classifier)      & 128 & 2048 & 1000 \\ 
5 & VGG-16 - FC (classifier)         & 64  & 4096 & 4096 \\ 
6 & MobileNetV2 - Bottleneck    & 1316 & 24  & 144   \\ 
\noalign{\hrule height 2pt} 
\end{tabular}%
}
\label{tab:gemm_dims}
\end{table}



\begin{table}[t]
\centering
\caption{Cost function weights for optimization templates.}
\resizebox{0.75\linewidth}{!}{%
\begin{tabular}{!{\vrule width 2pt}c!{\vrule width 2pt}c|c|c|c|c|c!{\vrule width 2pt}}
\noalign{\hrule height 2pt} 
 \textbf{Optimization template} & $\boldsymbol{\alpha}$ & $\boldsymbol{\beta}$& $\boldsymbol{\gamma}$ & $\boldsymbol{\theta}$  & $\boldsymbol{\zeta}$ & $\boldsymbol{\eta}$\\
\noalign{\hrule height 2pt} 
T1       & 1    &  1  & 1   & 1    & 1 & 1\\
T2        & 0.8  & 0.2 & 0.1 & 0.1  & 0.2 & 0.7 \\
T3    & 0.1  & 0.1 & 0.7 & 0.7  & 0.1 & 0.1 \\
T4     & 0.6  & 0.6 & 0.1 & 0.1  & 0.6 & 0.6\\
\noalign{\hrule height 2pt} 
\end{tabular}%
}
\label{tab:optimization_profile_table}
\end{table}

\subsection{Analysis of CarbonPATH PPAC models}
\label{sec:result-ppac}
In this section, we evaluate our modeling framework by analyzing the impact of architectural considerations and our modeling approach on the PPAC metrics of the HI system. We also perform a comparison against prior work~\cite{chiplet-gym}. Note that no prior work simultaneously considers PPAC and CFP for HI systems. So in this section of our results, we compare only our PPAC models against prior work, i.e., without CFP. In later sections, we analyze the impact of CFP. 

\subsubsection{Impact of number of chiplets on system latency}



\begin{figure}[t]
\centering
\includegraphics[width=0.95\linewidth]{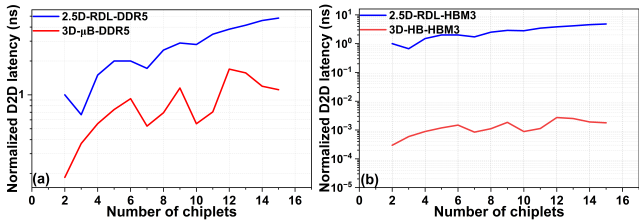}
\caption{Normalized D2D latency for different number of chiplets for WL1 for (a) \textit{2.5D-RDL-DDR5} and \textit{3D-\textmu B-DDR5} normalized to \textit{2.5D-RDL-DDR5} with 2 chiplets and (b) \textit{2.5D-RDL-HBM3} and \textit{3D-HB-HBM3} packaging and memory configurations normalized to \textit{2.5D-RDL-HBM3} with 2 chiplets.}
\label{fig:d2d_lat_vs_num_chiplet}
\end{figure}

\noindent
While chiplets reduce the cost and CFP of an HI system, they introduce performance overheads due to die-to-die (D2D) latency, particularly in 2.5D integration. To evaluate the impact of this latency, we design an experiment that varies the number of chiplets while keeping the chiplet configuration fixed at~\textit{128-7-1024}, the workload mapping fixed at~\textit{1-OS-0}, protocol fixed at \textit{UCS} protocol for 2.5D, \textit{UC3} for 3D, and show the trends for two different integration and packaging interconnects.
As an example, Fig.~\ref{fig:d2d_lat_vs_num_chiplet}(a) compares~\textit{2.5D-RDL-DDR5} with~\textit{3D-\textmu B-DDR5}, normalized to~\textit{2.5D-RDL-DDR5} with 2 chiplets and (b) compares~\textit{2.5D-RDL-HBM3} with~\textit{3D-HB-DDR5}, normalized to~\textit{2.5D-RDL-HBM3} with 2 chiplets\footnote{In the interest of page limitations, we show the results only for one specific workload, but our analysis for other workloads follow similar trends and can be found in our GitHub repository~\cite{carbonpath-gh}.} under different numbers of chiplets in the HI system.  In both cases, 3D integration achieves lower overall D2D latency than 2.5D integration due to the higher D2D bandwidth enabled by 3D packages, which benefit from shorter interconnects and larger numbers of I/Os distributed across the chip surface. 
As the number of chiplets increases, the total communication latency also increases because more data must be exchanged across chiplets, which becomes the bottleneck. 
This effect is pronounced during the reduction phase of the GEMM workload.  



We also observe a non-monotonic increase in D2D latency as the number of chiplets increases (for example, between six and seven chiplets). These fluctuations arise from topology-dependent D2D overheads for certain chiplet counts; the HI system interconnect topology requires extra hops and partial-result transfers to the destination chiplet during the reduction phase, increasing the effective D2D latency (see Sec~\ref{sec:model-latency}). These non-monotonic behaviors highlight the need to model D2D latency accurately. ChipletGym~\cite{chiplet-gym} assumes fixed D2D latencies of 17.2 ps for 2.5D and 1.6 ps for 3D, independent of the interconnect or topology or number or size of chiplets, whereas CarbonPATH models latency as a function of I/O count and package- and protocol-specific bandwidth, allowing it to capture the topology-dependent behavior.

\subsubsection{Impact of packaging tech. and interconnect on energy and cost}
\label{sec:energy-v-pkg}

\begin{figure}[t]
\centering
\includegraphics[width=0.95\linewidth]{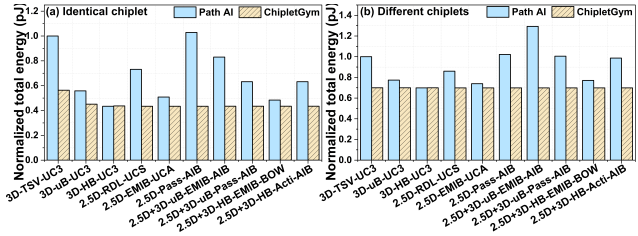}
\caption{Normalized energy variation with package-protocol combination for WL1 on a HI system with DDR5 DRAM and \textit{1-OS-0} workload mapping for a case with (a)~\textit{identical chiplet system}; and (b)~\textit{different chiplet system} normalized to \textit{3D-TSV-UC3}.}
\label{fig:energy-vs-pkg}
\end{figure}

\begin{figure}[t]
\centering
\includegraphics[width=0.95\linewidth]{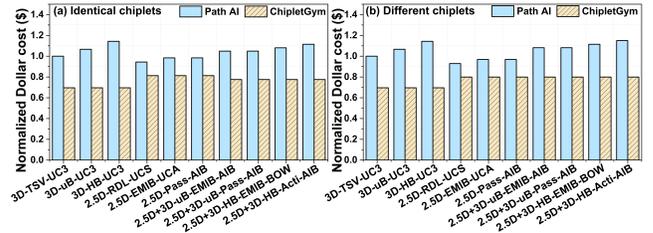}
\caption{Normalized cost variation with package-protocol combination for WL1 on a HI system with DDR5 DRAM and \textit{1-OS-0} workload mapping for a case with (a)~\textit{identical chiplet system}; and (b)~\textit{different chiplet system}  normalized to \textit{3D-TSV-UC3}.}
\label{fig:cost-vs-pkg}
\end{figure}

\noindent
With changes in packaging integration and interconnect types, the energy required to perform computation and access DRAM varies due to differences in the energy per bit of data transfer across the dies. Similarly, dollar cost varies due to differences in manufacturing processes and bonding yields across interconnect types.  To assess the impact, we conduct an experiment in which we vary the package and protocol for 2.5D, 3D, and 2.5D+3D integration while holding other design considerations constant. 


Fig.~\ref{fig:energy-vs-pkg}(a) and (b) show the energy for the~\textit{identical chiplet system} and~\textit{different chiplet system}, respectively, normalized to \textit{3D-TSV-UC3}. Fig.~\ref{fig:cost-vs-pkg}(a) and (b) show the cost for the~\textit{identical chiplet system} and~\textit{different chiplet system}, respectively, normalized to \textit{3D-TSV-UC3}.In both the experiments, we maintain a fixed workload mapping of \textit{1-OS-0} and utilize DDR5 memory.
For total energy (Fig.~\ref{fig:energy-vs-pkg}), in both cases (identical and different sized chiplets) the 3D hybrid-bonding with the UCIe 3D protocol has the least energy, as it has a faster memory access via the low-pitch hybrid bonding with high bandwidth and therefore lower energy. For the identical-chiplet case, \textit{2.5D-Pass-AIB} has the highest energy for this workload, due to the large latencies of 2.5D integration. For the case with different chiplets, we observe that the \textit{2.5D+3D-uB-EMIB-AIB} has the highest energy, because the EMIB reduces bandwidth relative to other interconnect types and therefore increases latency and energy.   
For dollar cost, in both cases (identical and different-sized chiplets), the \textit{2.5D-RDL-UCS} has the least cost as this is the most mature integration technology and interconnect type with the highest yield. Since 3D hybrid bonds have the lowest bonding yield in both cases, they incur the highest cost, whereas TSVs are the least expensive interconnect type for 3D integration in both cases.

We also compare the total energy and cost against ChipletGym~\cite{chiplet-gym} in the same figures. The ChipletGym energy model does not account for protocol overheads or SRAM energy and instead relies only on energy per MAC operation. In contrast, CarbonPATH includes DRAM, SRAM, compute, and die-to-die energy in its calculations. As a result, CarbonPATH reports higher total energy, as shown in the figure, reflecting a more comprehensive and realistic system-level energy model. The ChipletGym cost model assumes a constant bonding yield of 0.99 and does not account for differences in bonding yield across packaging types; as a result, it reports lower cost values than CarbonPATH.

\subsubsection{Impact of packaging tech. and interconnect on latency and cost}
\label{sec:lat-v-cost-scatter}

\noindent
In this experiment, we highlight the tradeoff between latency and cost. We vary the package and protocol combination for \textit{different chiplet system} 
with DDR5 memory and~\textit{1-OS-0} workload mapping. We explore all combinations of protocol, package, for all three integration types: 2.5D, 3D, and 2.5D+3D cases for WL1. 

\begin{figure}[t]
\centering
\includegraphics[width=\linewidth]{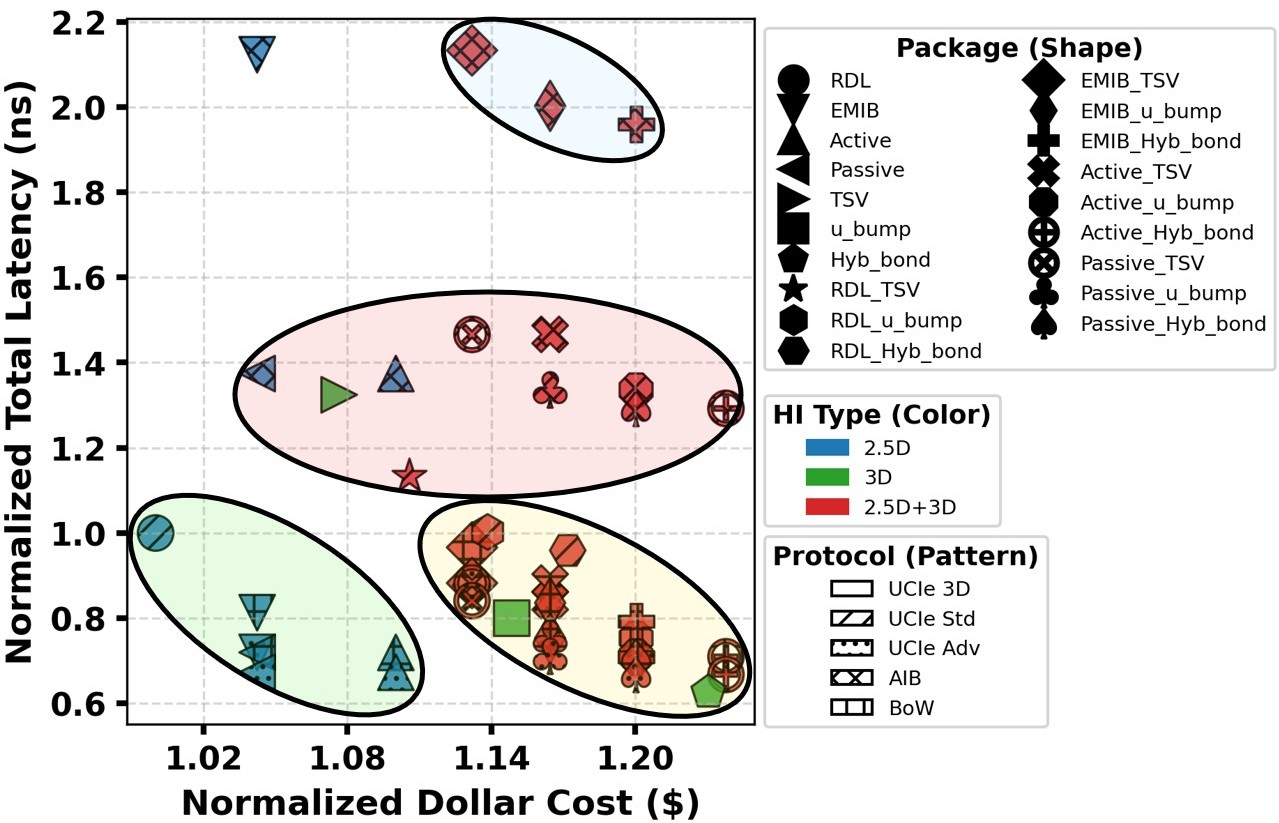}
\caption{Scatter plot that compares normalized latency vs normalized dollar cost for pkg-protocol combinations for WL1 run on \textit{different chiplet system} configuration and normalized to \textit{2.5D-RDL-DDR5} with \textit{UCS}.}
\label{fig:scatter-latency-vs-cost}
\end{figure}

The scatter plot in Fig.~\ref{fig:scatter-latency-vs-cost} shows the variation in total latency versus dollar cost for all 43 interconnect and protocol pairs normalized to \textit{2.5D-RDL-DDR5} with \textit{UCS}. Compared to 2.5D (blue scatter points), 3D packages (green points) provide higher I/O density and more bumps per mm$^2$, enabling vertical connections across the entire die area. This results in substantially higher die-to-die bandwidth and lower latency. These gains, however, come at the expense of manufacturing complexity, as 3D integration typically exhibits lower yield and higher cost than 2.5D. The 2.5D integration points lie on the left side of the figure, 
highlighted in the green region on the scatter plot, while the 3D points are spread across both axes in the scatter plot. The yellow region achieves the lowest latency, utilizing 3D and 2.5D+3D packaging, while the high bandwidth of these architectures optimizes performance, it comes with a higher cost. The red region occupies the center of the plot and is dominated by 2.5D+3D configurations that offer intermediate values for both latency and monetary cost. Finally, the blue region comprises EMIB-based designs, which exhibit higher delays for this specific workload, placing them in the high-latency sector of the graph.
Across all cases, we observe a nearly 10$\times$ variation in latency between the minimum latency and maximum latency scatter points, highlighting the breadth of the design space and the significant impact of packaging choice on system latency.

\subsubsection{Impact of workload mapping on latency}
One of the key differences between CarbonPATH and ChipletGym is that CarbonPATH models workload mapping on the specific HI system architecture.  We set up an experiment that varies the workload mapping parameters to measure the impact of different workload mapping strategies on latency. 

\begin{figure}[t]
\centering
\includegraphics[width=0.95\linewidth]{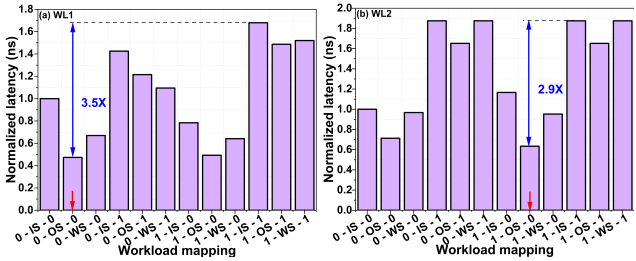}
\caption{Normalized latency under different mappings for 2.5D+3D HI system with \textit{64-7-256, 96-7-512, 128-7-1024}, and \textit{192-7-2048} normalized to \textit{0-IS-0} configuration for (a) WL1 and (b) WL2.}
\label{fig:wl_map_1p5}
\end{figure}

As a representative example, Fig.~\ref{fig:wl_map_1p5} illustrates the performance variation of latency with different workload mapping configurations shown along the x-axis for a hybrid 2.5D+3D integrated HI system with~\textit{96-7-512, 128-7-1024} stacked in 3D configuration, and~\textit{64-7-256, 192-7-2048} connected with~\textit{128-7-2048}. All values normalized to~\textit{0-IS-0} workload mapping configuration. The stacked chiplets use~\textit{3D-HB-DDR5} in \textit{UC3}, and the 2.5D chiplets use \textit{2.5D-RDL-DDR5} in \textit{UCS}\footnote{The 2.5D+3D HI architecture is represented by the notation \textit{2.5D-RDL-3D-HB-DDR5}.}. 
Fig.~\ref{fig:wl_map_1p5}(a) illustrates the results for WL1, while Fig.~\ref{fig:wl_map_1p5}(b) presents those for WL2. Among all evaluated configurations, the WL2 combination labeled~\textit{1-OS-0} (highlighted in red arrow) yields the best performance for this architecture, indicating that assigning workloads in ascending order, disabling K-dimension splitting, and applying an output-stationary dataflow is the most effective setup for this workload and architecture. For WL1, the mapping style~\textit{0-OS-0} yielded the best latency. We also observe 3.5X and 2.9X variations in latency between the minimum and maximum values across the different mapping styles for these two workloads. 

We observe the OS dataflow to be the lowest in latency for both workloads because, in the OS dataflow, the partial sums remain local to each compute core, reducing traffic across the dies, which can be helpful when memory and D2D bandwidths are bottlenecks. The assignment order differs between WL1 and WL2, highlighting that workload characteristics and HI system heterogeneity can vary with assignment order. WL2 benefits from a smallest-first ordering to reduce fragmentation and better match resource granularity, while WL1 benefits from a largest-first ordering and keeps the critical path on higher-capability cores. These results show that workload size, HI system characteristics, and the assignment order jointly influence the best mapping choice.

\subsection{Analysis of CarbonPATH tradeoffs between PPAC and CFP}
\label{sec:carbon-results}
In this section, we examine the importance of the carbon model for chiplet-based HI systems using CarbonPATH to illustrate its effect on system-level results.

\subsubsection{Impact of number of chiplets on Perf-SI}

\begin{figure}[t]
\centering
\includegraphics[width=0.95\linewidth]{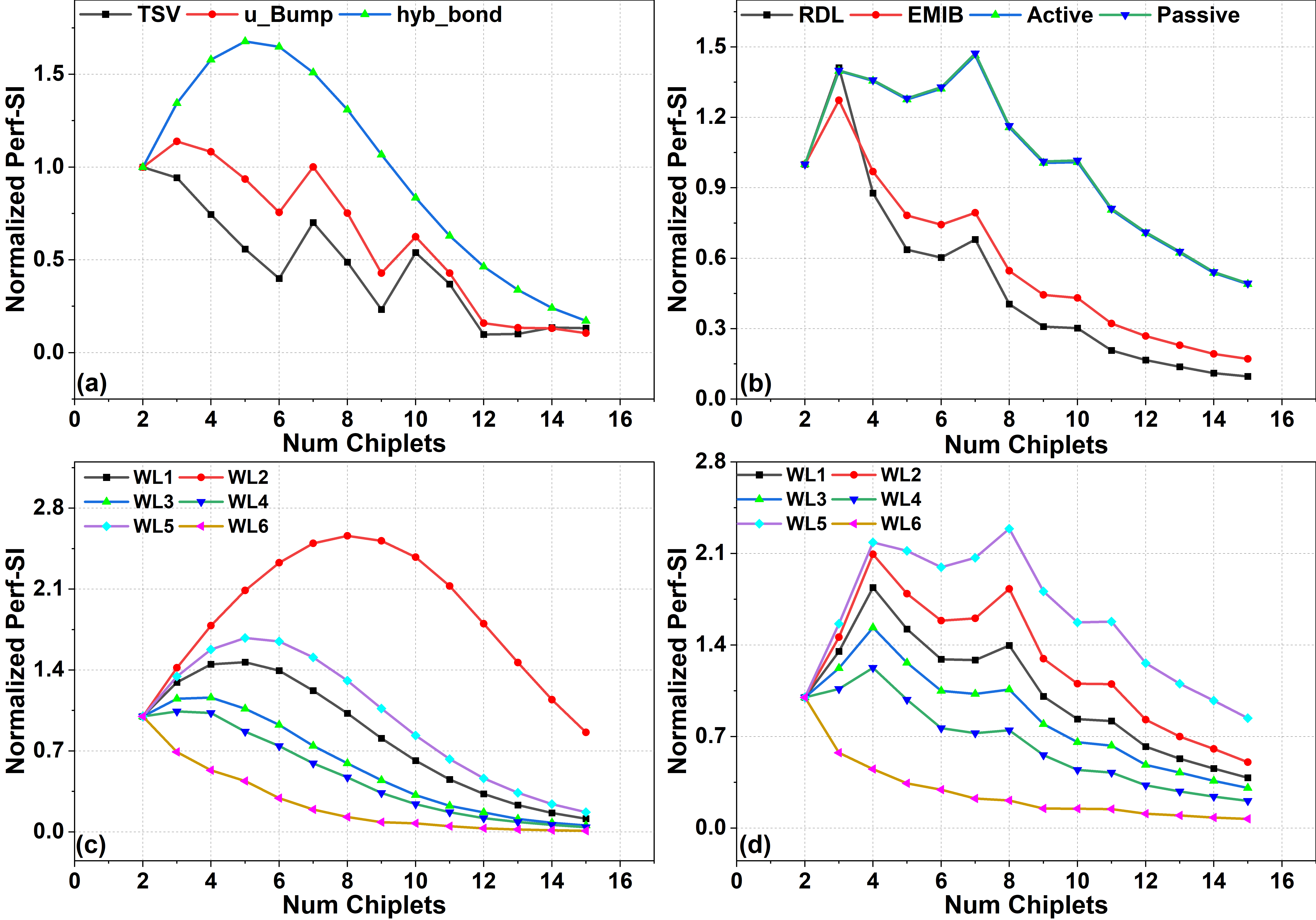}
\caption{Variation in normalized Perf-SI across different number of \textit{127-7-1024} chiplets for (a) WL1 and different 3D packaging interconnects with DDR5, (b) WL1 for different 2.5D packaging interconnect with DDR5, (c) \textit{3D-HB-DDR5} for all workloads, and (d) \textit{2.5D-Acti-DDR5} for all workloads, normalized to the value of a 2-chiplet system. }
\label{fig:carbon-perf-vs-num-chiplet}
\end{figure}

HI can reduce overall CFP and cost\cite{eco-chip,3d-carbon}, but the performance implications must be quantified alongside these gains. Fig.~\ref{fig:carbon-perf-vs-num-chiplet} plots Perf-SI (as defined in Sec~\ref{sec:background}), normalized to the two-chiplet baseline. We utilize the Perf-SI metric to evaluate carbon efficiency~\cite{mobile-CFP}. Higher value indicates superior HI configurations that maximize computational throughput while minimizing carbon emissions.
Increasing the number of chiplets enhances the system's compute capabilities but adds penalties in terms of system area and communication latency. This creates a distinct inflection point in performance: throughput initially improves, but eventually declines as D2D and memory communication overheads dominate. In contrast, the environmental cost trends upward. The total CFP increases primarily due to the rising embodied CFP associated with more chiplets, whereas operational CFP varies with energy and latency. However, compared to a monolithic system, chiplet-based systems are more sustainable~\cite{eco-chip}. 

With the chiplet configuration of each chiplet fixed at \textit{128-7-1024}, with DDR5 memory and workload mapping \textit{0-OS-1}, Fig.~\ref{fig:carbon-perf-vs-num-chiplet}(a) compares the Perf-SI across all 3D packaging interconnects in \textit{UC3} for WL1, (b) compares all 2.5D packaging interconnect types in \textit{UCS}, (c) shows variation across workloads for a fixed 3D hybrid-bond package in \textit{UC3}, and (d) does the same for a 2.5D active interposer in \textit{UCS}. It can be seen that for both 2.5D and 3D packages, higher die-to-die bandwidth packages yield larger gains as chiplet count increases—peaking at five chiplets for 3D hybrid bonding and seven chiplets for 2.5D active/passive interposers from (a) and (b), respectively. This is because higher-bandwidth packages reduce overall D2D communication time and support increased traffic and partial-sum exchange; however, these benefits come with packaging overheads, higher monetary cost, and lower yield. The monotonic Perf-SI across the number of chiplets is due to our accurate topology-aware D2D model, which accounts for routing, hops, and partial-sum transfers.

Fig.~\ref{fig:carbon-perf-vs-num-chiplet}(c) and (d) highlight the workload sensitivity for 3D hybrid bonding and 2.5D active interposer. Profiling these variations across different workloads and packaging protocols is essential for identifying configurations that maximize Perf-SI. The results demonstrate that for large workloads (WL1, WL2, WL5), increasing the number of chiplets initially improves Perf-SI. This improvement arises from the reduction in CFP relative to the 2-chiplet baseline, as greater disaggregation improves the yield of individual dies and thereby lowers overall embodied CFP. In 3D scenarios, performance also improves with increased stacking, owing to the higher available bandwidth enabled by vertical interconnects. However, beyond a certain threshold in chiplet count, Perf-SI begins to decline, as diminishing performance gains and rising overheads offset the benefits of improved yield. In contrast, for smaller workloads (WL6), a larger number of chiplets is not beneficial for Perf-SI. In these cases, communication overheads quickly dominate, degrading performance and increasing CFP. These findings highlight the importance of carefully tuning the chiplet count to the specific package–workload combination in order to achieve optimal Perf-SI.


\subsubsection{Impact of packaging tech. and interconnect on Perf-SI and cost}

\begin{figure}[t]
\centering
\includegraphics[width=0.95\linewidth]{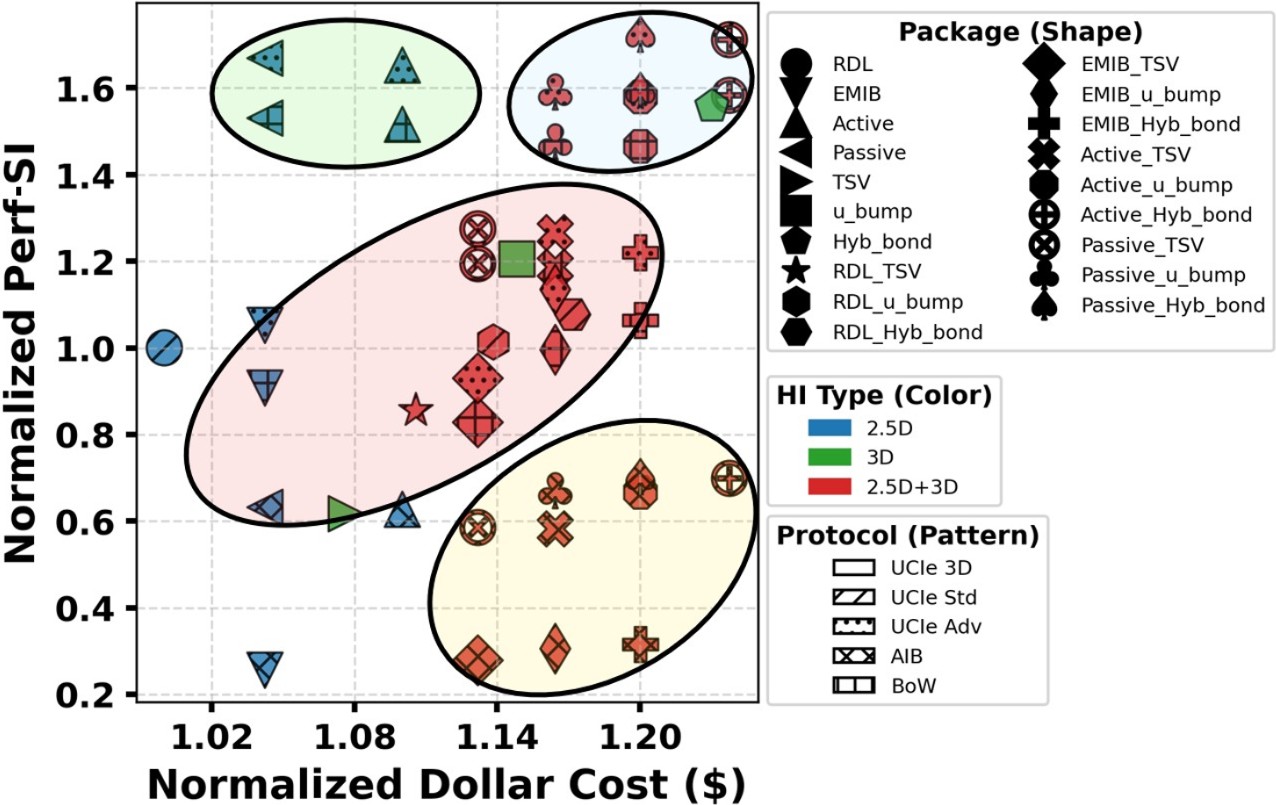}
\caption{Scatter plot of normalized Perf-SI  with normalized monetary cost for package–protocol combinations under WL1 on the HI system, shown for ~\textit{different chiplet system} normalized to \textit{2.5D-RDL-DDR5} \textit{UCS}.}
\label{fig:carbon-scatter-plot}
\end{figure}

Fig.~\ref{fig:carbon-scatter-plot} shows the scatter plot of Perf-SI with dollar cost for all possible 43 package-protocol combinations, normalized to \textit{2.5D-RDL-DDR5} \textit{UCS} for a \textit{different chiplet system} with DDR5 and \textit{0-OS-1} workload mapping. This figure is similar to Fig.~\ref{fig:scatter-latency-vs-cost} but now includes a sustainability dimension. The data shows a broad distribution across cost and Perf-SI axes.

We see varying Perf-SI values even at the same dollar cost, and the most expensive options are not necessarily the most efficient from a carbon perspective. The region in the top-left, highlighted by a green circle, is the best in the plot, with the highest Perf-SI at the lowest cost. Several 2.5D HI options (active/passive interposer with UCIe-Advance or BoW) are clustered in this top-left region. These 2.5D advanced packages are less expensive than 3D packages and deliver good performance for this particular WL1. The pink-highlighted region shows a diverse mix of HI types, exhibiting a broad distribution along both the cost and Perf-SI axes. In this region, 2.5D configurations cluster at the lower end of the cost spectrum, reflecting the relative maturity and affordability of 2.5D packaging. In contrast, the 3D and hybrid 2.5D+3D cases are concentrated in the higher-cost region, a direct consequence of the elevated manufacturing complexity and expense associated with vertical stacking technologies.
The yellow region highlights 2.5D+3D cases, which are characterized by high implementation costs (due to 3D packaging) and low Perf-SI caused by poor yields. 
The region highlighted in blue on the top right are cases that are 2.5D+3D and 3D with micro bumps and hybrid bonds, that have good bandwidth between dies and generally high performance, but these come with a higher cost. This scatter plot is a practical tool for architectural decisions: selecting the final package–protocol combination should weigh both dollar cost and Perf-SI. The picture shifts with workload, and system configuration (identical or different chiplets, and chiplet count). This visualization is a useful tool for design space exploration. By applying specific constraints on cost, performance, or sustainability, architects can identify the feasible region—isolating valid system configurations while filtering out options that violate target specifications.

\subsubsection{Impact of workload mapping on Perf-SI}

\begin{figure}[t]
\centering
\includegraphics[width=0.95\linewidth]{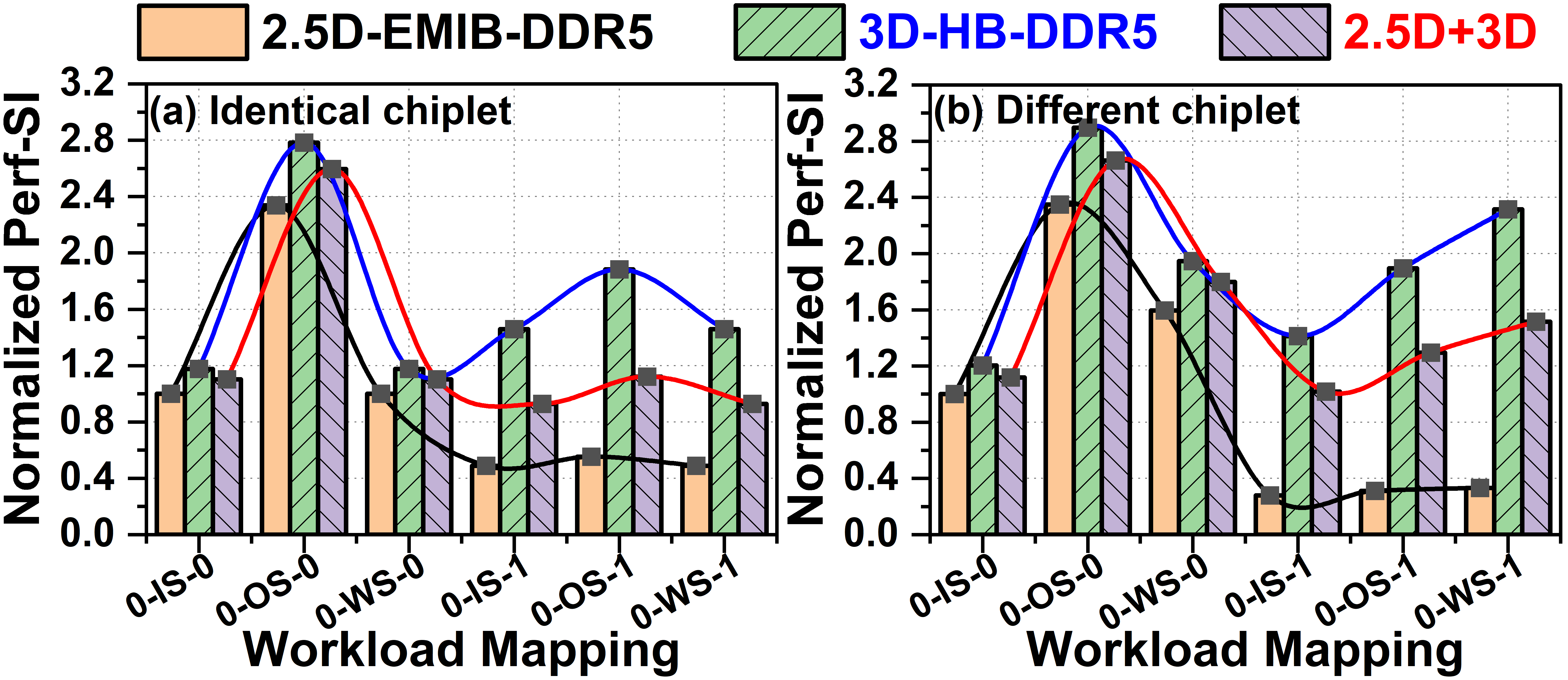}
\caption{Variation in Perf-SI for \textit{2.5D-EMIB-DDR5} with \textit{UCS}, \textit{3D-HB-DDR5} with \textit{UC3}, and 2.5D+3D (\textit{2.5D-EMIB-3D-HB-DDR5}) HI-pkg types for (a)~\textit{identical chiplet system}, and (b)~\textit{different chiplet system} configuration run for WL1, each of them normalized to \textit{2.5D-EMIB-DDR5} with \textit{0-IS-0}. } 
\label{fig:carbon-wl_map}
\end{figure}

Workload mapping plays a critical role in fully utilizing chiplets and memory in a heterogeneous system. Fig.~\ref{fig:carbon-wl_map} reports normalized Perf-SI values, each normalized to Perf-SI of \textit{2.5D-EMIB-DDR5} in \textit{UCS} protocol with \textit{0-IS-0} workload mapping. Across all points in the plot, the underlying system design remains constant—meaning the embodied CFP is identical for every configuration. We also fix the memory type to DDR5, and use \textit{3D-HB-DDR5} with \textit{UC3} or \textit{2.5D-EMIB-DDR5} with \textit{UCS} for the corresponding 3D, 2.5D, and 2.5D+3D integration. Fig.~\ref{fig:carbon-wl_map}(a) shows for~\textit{identical chiplet system} configuration and Fig.~\ref{fig:carbon-wl_map}(b) shows for~\textit{different chiplet system} configuration. In both plots, 3D packaging achieves the best Perf-SI values, followed by the 2.5D+3D package. It can be observed that enabling~\textit{split-K} has a highly asymmetric effect across HI types. For 2.5D,~\textit{split‑K} reduces performance below the baseline for all dataflows, indicating that the additional partial‑sum traffic is bottlenecked by the limited 2.5D interposer bandwidth. For the 3D and 2.5D+3D integrations, it can be seen that~\textit{split-K} leverages the high D2D bandwidth, thereby reducing overall execution time, lowering operational CFP, and improving perf-SI. In configurations where~\textit{split-K} is disabled, the OS  dataflow has the highest Perf-SI across all HI-types. This advantage arises because in the OS dataflow, the partial sum is maintained locally, thereby minimizing the data-movement overheads typically associated with IS and WS dataflows. Our results demonstrate that 3D packaging is the superior architecture for parallelizing operations via split-K. Ultimately, the optimal dataflow strategy relies on a complex interplay between integration type, packaging interconnect, and workload characteristics. CarbonPATH models these workload-mapping choices and captures their system-level impact on performance and CFP.

\subsubsection{Impact of packaging on CFP and cost}

\begin{figure}[t]
\centering
\includegraphics[width=0.95\linewidth]{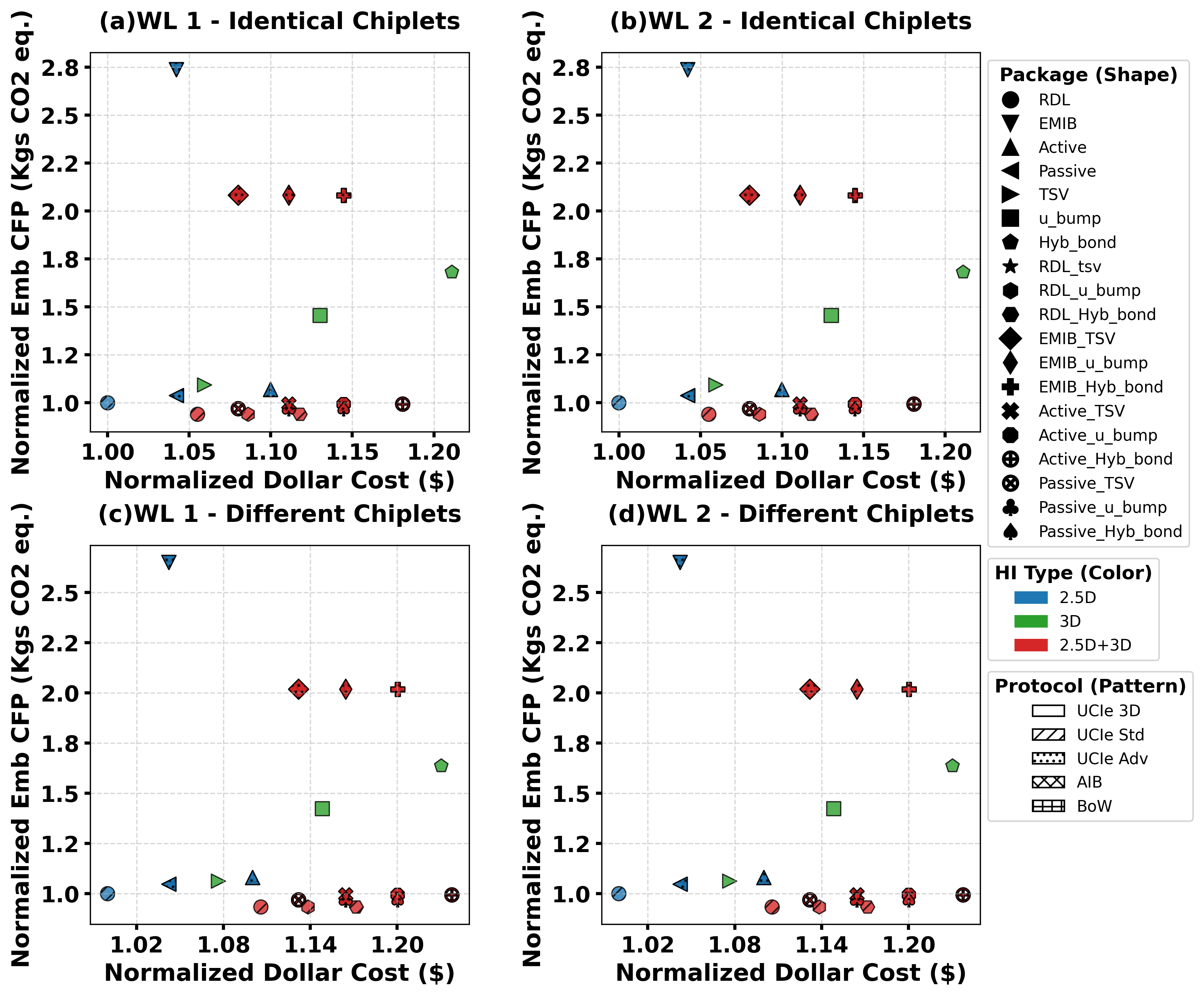}
\caption{Variation in normalized embodied carbon with normalized cost for package-protocol combinations for (a) WL1 ~\textit{identical chiplet system}, (b) WL2 ~\textit{identical chiplet system}, (c) WL1 ~\textit{different chiplet system}, and (d) WL2 ~\textit{different chiplet system}, all values normalized to ~\textit{2.5D-RDL-DDR5} with \textit{UCS} protocol.}
\label{fig:carbon-vs-cost}
\end{figure}

There is prior literature that often treats monetary cost as a proxy for carbon emissions~\cite{cost-carbon-proxy}, Fig.~\ref{fig:carbon-vs-cost} demonstrates that these metrics are not directly correlated. The figure plots the normalized embodied CFP against the normalized monetary cost. Fig.~\ref{fig:carbon-vs-cost}(a) and (b) illustrate the variation for~\textit{identical chiplet system} configuration for workloads WL1 and WL2, respectively. Fig.~\ref{fig:carbon-vs-cost}(c) and (d) show the results for WL1 and WL2 for \textit{different chiplet system} configuration, with all values normalized to ~\textit{2.5D-RDL-DDR5} with \textit{UCS} protocol. We fix the workload mapping at \textit{0-OS-1} and utilize DDR5 memory. The plots clearly demonstrate that there is no direct or linear relationship between cost and embodied CFP. Designs utilizing the EMIB package exhibit a significantly higher total CFP. This is attributed to the EMIB’s dense silicon bridge, which features approximately 250 wires per mm and fine metal routing layers between the dies.
2.5D packages (blue) occupy the lower end of the cost spectrum compared to 3D but do not consistently yield the lowest total CFP. The 3D packages (green) tend to fall within the mid-to-high CFP range, spanning the entire cost spectrum. The wide distribution of data points indicates that no single package-protocol combination is optimal across all metrics. Ultimately, both cost and CFP are influenced by a complex combination of packaging, protocols, bonding yields, workloads, mapping styles, and chiplet configurations. 


\subsection{CarbonPATH SA optimization results}
\label{sec:result-sa-overall}
The CarbonPATH framework that leverages the developed models and a SA engine is employed to determine the optimal HI system for various workloads (Table~\ref{tab:gemm_dims}) under the distinct optimization templates outlined in Table~\ref{tab:optimization_profile_table}. To evaluate the CarbonPATH optimization engine, we compare its outputs with those of three flows. The first is an SA-based optimization that uses models from ChipletGym. The second is an SA-based engine that uses CarbonPATH models but with $\zeta= 0$, and $\eta=0$ (to highlight the importance of CFP and since ChipletGym does not account for CFP, we call this CarbonPATH w/o carbon), and the third is the proposed CarbonPATH engine that uses all CarbonPATH models. For each of the three flows, we tune the SA hyperparameters and calibration values to ensure convergence to the best possible solution.  In the rest of this section, we compare the results generated by these three optimization flows. The first subsection compares the PPAC and CFP metrics of the generated solutions, while the next subsection compares the converged solution. 

\subsubsection{Comparison of generated solution's PPAC and CFP metrics}
\label{sec:result-sa-compare}
\begin{figure}[t]
\centering
\includegraphics[width=0.95\linewidth]{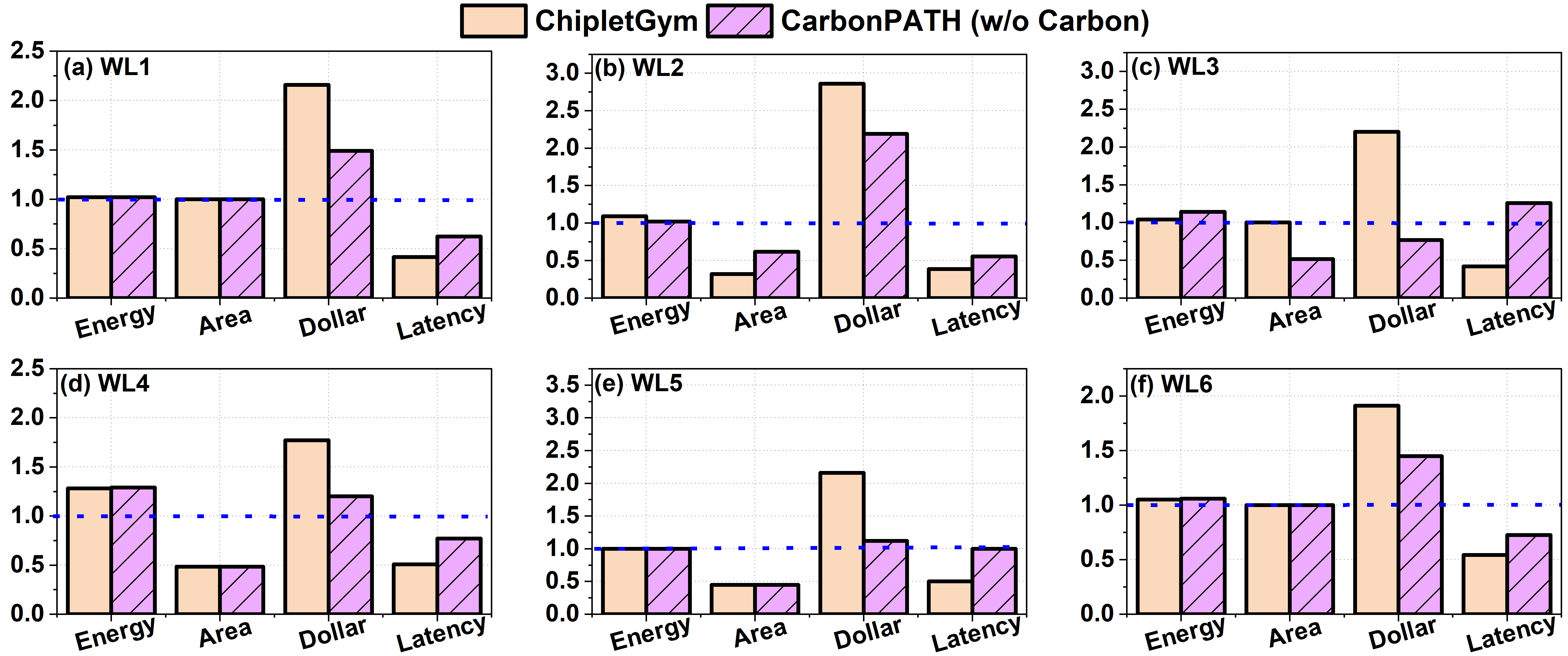}
\caption{CarbonPATH's normalized area, energy, latency, dollar cost, and SA-Cost comparison with Chiplet-Gym for all WLs running in T1 profile.}
\label{fig:sa_barplot_2p1}
\end{figure}

\begin{figure}[t]
\centering
\includegraphics[width=0.95\linewidth]{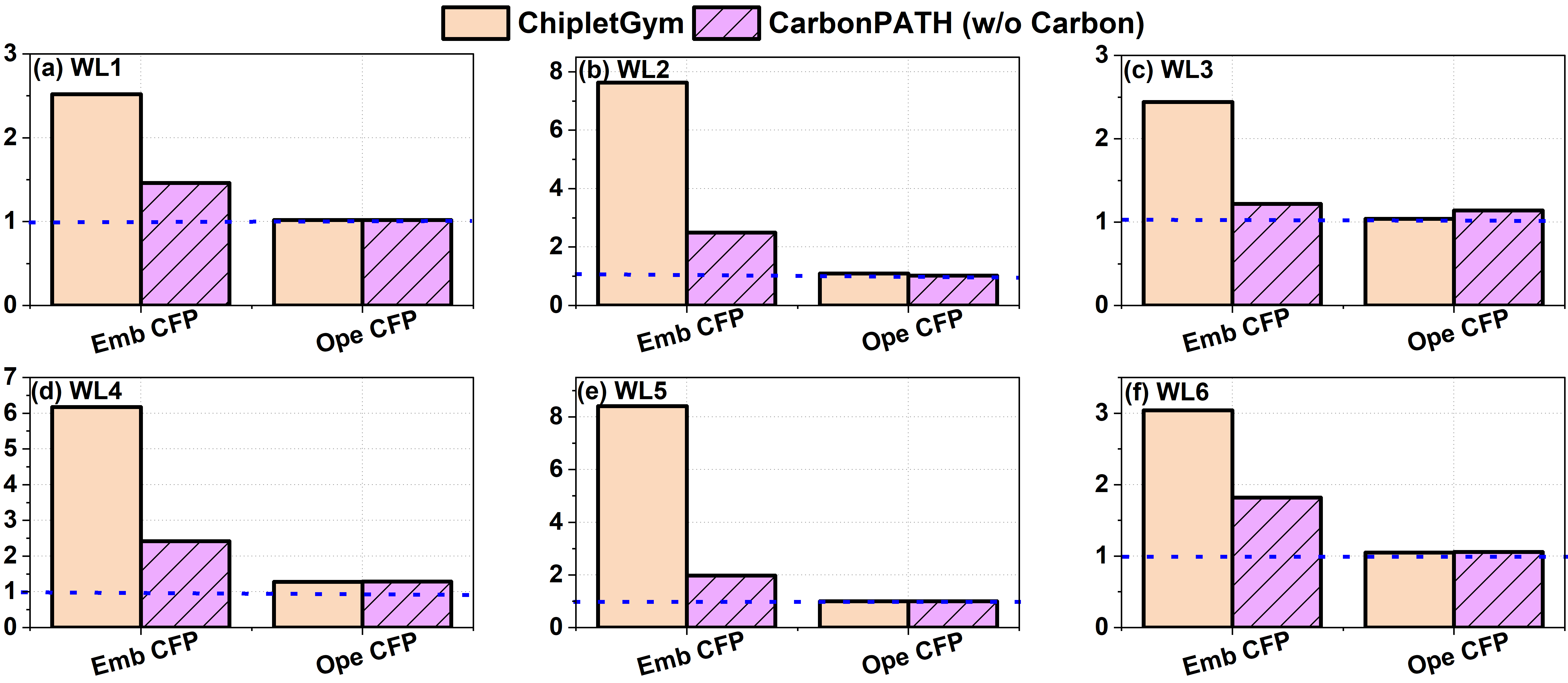}
\caption{CarbonPATH's normalized embodied and operational CFP comparison with Chiplet-Gym and CarbonPATH w/o carbon for all WLs running in T1 profile.}
\label{fig:sa_barplot_2p1_carbon}
\end{figure}


Fig.~\ref{fig:sa_barplot_2p1} compares the energy, latency, monetary cost, and area for all workloads for template T1 for the three different flows. The results are normalized to CarbonPATH (blue dotted line). The values greater than 1 indicate worse metrics. Across all workloads, the metrics for CarbonPATH is the lowest and equal to CarbonPATH w/o carbon compared to the ChipletGym optimization flow, demonstrating differences between modeling approaches of ChipletGym and CarbonPATH. The only exception is latency, where ChipletGym reports lower values. This discrepancy arises from modeling assumptions: whereas ChipletGym assumes a fixed die-to-die latency, CarbonPATH models latency as a function of packaging technology, offering a more accurate representation of system behavior.

\noindent
Fig.~\ref{fig:sa_barplot_2p1_carbon} illustrates embodied and operational CFP for all workloads under profile T1 of the generated solutions from the three optimization flows. The plot demonstrates that CarbonPATH consistently achieves lower CFP across all workloads, indicating its effectiveness in identifying the optimal HI system when sustainability is prioritized, since the other optimization flows do not consider CFP in their cost functions. 
Quantitatively, enabling sustainability optimization in CarbonPATH delivered a 1.9X average embodied CFP improvement, with gains scaling from 1.9X for the T1 profile to a peak of 3.16X for the T4 profile, compared to CarbonPATH w/o carbon.


\begin{table}[ht]
\caption{Comparing the CFP and PPAC of the solutions generated by the three optimization flows for all WLs and templates.}
\centering
\resizebox{\linewidth}{!}{%
\begin{tabular}{!{\vrule width 2pt}l!{\vrule width 2pt}cccc|cccccc!{\vrule width 2pt}}
\noalign{\hrule height 2pt}
 & \multicolumn{4}{c|}{\textbf{ChipletGym}} 
 & \multicolumn{6}{c!{\vrule width 2pt}}{\textbf{CarbonPATH w/o carbon}} \\
\noalign{\hrule height 1pt}
\textbf{WL-profile} 
  & \textbf{Energy} & \textbf{Area} & \textbf{Dollar} & \textbf{Latency} 
  & \textbf{Energy} & \textbf{Area} & \textbf{Dollar} & \textbf{Latency} & \textbf{Emb CFP} & \textbf{Ope CFP}  \\
\noalign{\hrule height 2pt}

WL1-T1 & 1.02  & 1	   & 2.16  & 0.416 & 1.02	& 1     & 1.49	& 0.624	& 1.46	& 1.02  \\
WL1-T2 & 0.929 & 0.894 & 3.594 & 0.313 & 0.992  & 0.463 & 1.552 & 0.834 & 2.039 & 0.992  \\
WL1-T3 & 1.001 & 0.473 & 1.859 & 0.667 & 1      & 1     & 1     & 1     & 1     & 1      \\ 
WL1-T4 & 0.798 & 1.794 & 4.507 & 0.245 & 0.862  & 0.929 & 2.671 & 0.473 & 4.342 & 0.862  \\ \hline

WL2-T1 & 1.09  & 0.321 & 2.86  & 0.389 & 1.02  & 0.619  & 2.19  & 0.556 & 2.5   & 1.02   \\
WL2-T2 & 1.014 & 1     & 2.812 & 0.350 & 1.004 & 1      & 1.453 & 0.675 & 1.763 & 1.004  \\
WL2-T3 & 1.011 & 0.293 & 2.364 & 0.518 & 0.966 & 1.198  & 1.276 & 0.860 & 1.091 & 0.967  \\
WL2-T4 & 0.839 & 2.070 & 5.841 & 0.150 & 0.828 & 2.070  & 2.077 & 0.429 & 1.517 & 0.828  \\ \hline

WL3-T1 & 1.04  & 1	    & 2.2	& 0.421	& 1.14	& 0.518 & 0.768	& 1.26	& 1.22	& 1.14   \\
WL3-T2 & 0.899 & 1.932  & 3.687 & 0.251 & 1     & 1     & 1     & 1     & 1     & 1      \\
WL3-T3 & 1.001 & 0.473  & 1.859 & 0.666 & 0.962 & 0.764 & 0.649 & 1.583 & 0.700 & 0.962  \\
WL3-T4 & 0.462 & 1      & 4.456 & 0.190 & 0.668 & 0.250 & 1.794 & 0.638 & 6.610 & 0.668  \\ \hline

WL4-T1	& 1.28	& 0.483	 & 1.77	& 0.507	& 1.29	& 0.483	& 1.2	& 0.772	& 2.42	& 1.29   \\
WL4-T2  & 1.28  & 0.483  & 1.77 & 0.507 & 1.29  & 0.483 & 1.2   & 0.772 & 2.42  & 1.29   \\ 
WL4-T3  & 1.09  & 0.619  & 1.97 & 0.728 & 1     & 1     & 1     & 1     & 1     & 1      \\
WL4-T4  & 0.82  & 1.79   & 5.11 & 2.81  & 1.01  & 0.449 & 1.84  & 0.682 & 2.47  & 1.01   \\ \hline

WL5-T1	& 1	    & 0.449	& 2.16	& 0.5	& 1	    & 0.449	& 1.12 &	1	& 1.98 &	1    \\
WL5-T2  & 1.007 & 3.602 & 2.534 & 0.344 & 1     & 1 & 1     & 1     & 1     & 1          \\
WL5-T3  & 1.002 & 0.224 & 1.725 & 0.581 & 1     & 1 & 1     & 1     & 1     & 1          \\
WL5-T4  & 1.009 & 1     & 2.812 & 0.344 & 1.001 & 1 & 1.906 & 0.500 & 3.043 & 1.001      \\ \hline

WL6-T1	& 1.05	& 1	 & 1.91	& 0.542 &	1.06	 & 1 & 	1.45 &	0.725 &	1.82	 & 1.06  \\
WL6-T2  & 1.053 & 1     & 1.906 & 0.542 & 1 & 1     & 1     & 1 & 1     & 1              \\
WL6-T3  & 0.999 & 0.449 & 1.466 & 0.748 & 1 & 1.615 & 0.924 & 1 & 1.228 & 1              \\
WL6-T4  & 1.053 & 1     & 1.906 & 0.542 & 1 & 1     & 1     & 1 & 1     & 1              \\ \hline

\noalign{\hrule height 2pt}
Average & 0.989 & 1.014 &	2.718 &	0.552 &	1.004 &	0.887 &	1.356 &	0.849 &	1.900 &	1.004 \\ \hline
\noalign{\hrule height 2pt}
\end{tabular}%
}
\label{tab:SA-numbers-compare-all}
\end{table}

\begin{table}[ht]
\centering
\caption{Comparison of optimized HI system for template T1.}
\resizebox{\linewidth}{!}{%
\setlength{\arrayrulewidth}{1pt}
\begin{tabular}{!{\vrule width 2pt} l !{\vrule width 2pt} c|c|c !{\vrule width 2pt} c|c|c !{\vrule width 2pt}}
\noalign{\hrule height 2pt}

& \multicolumn{3}{c!{\vrule width 2pt}}{\textbf{WL1 - T1 profile}}
& \multicolumn{3}{c!{\vrule width 2pt}}{\textbf{WL2 - T1 profile}} \\
\noalign{\hrule height 2pt}
\textbf{Param} & \textbf{ChipletGym} & \textbf{CarbonPATH w/o C} & \textbf{CarbonPATH}
                   & \textbf{ChipletGym} & \textbf{CarbonPATH wo C} & \textbf{CarbonPATH} \\
\hline
\textbf{\#chiplets}        & 3 & 2 & 2  & 6 & 3 & 2 \\
\textbf{\textit{A-T-S}}    & \textit{128-7-1024} X3 & \textit{128-7-1024} X2 & \textit{64-10-256, 128-7-1024}
                          & \textit{96-7-512} X6 & \textit{128-7-1024} x2, \textit{64-7-256} & \textit{64-7-256, 128-7-1024} \\
\textbf{\textit{I-P-M}}            & \textit{3D-HB-HBM3} & \textit{3D-HB-HBM3} & \textit{3D-HB-HBM3}
                          & \textit{3D-HB-HBM3} & \textit{3D-HB-HBM3} & \textit{2.5D-RDL-HBM3} \\
\textbf{Protocol}          & UCIe 3D & UCIe 3D & UCIe 3D
                          & UCIe 3D & UCIe 3D & UCIe Std \\
\textbf{\textit{O-D-K}}            & \textit{0-OS-0} & \textit{0-OS-0} & \textit{0-OS-1}
                          & \textit{0-OS-0} & \textit{0-OS-1} & \textit{0-OS-1} \\
\noalign{\hrule height 2pt}

& \multicolumn{3}{c!{\vrule width 2pt}}{\textbf{WL3 - T1 profile}}
& \multicolumn{3}{c!{\vrule width 2pt}}{\textbf{WL4 - T1 profile}} \\
\noalign{\hrule height 2pt}
\textbf{\#chiplets}        & 3 & 3 & 2  & 6 & 4 & 2 \\
\textbf{\textit{A-T-S}}    & \textit{128-7-1024} X3 & \textit{64-7-256} x2, \textit{96-7-512} & \textit{64-7-256, 128-7-1024}
                          & \textit{64-7-256} x6 & \textit{64-7-256} x4 & \textit{64-7-256, 96-7-512} \\
\textbf{\textit{I-P-M}}            & \textit{3D-HB-HBM3} & \textit{3D-HB-HBM3} & \textit{3D-HB-HBM3}
                          & \textit{3D-HB-HBM3} & \textit{3D-HB-HBM3} & \textit{3D-HB-HBM3} \\
\textbf{Protocol}          & UCIe 3D & UCIe 3D & UCIe 3D
                          & UCIe 3D & UCIe 3D & UCIe 3D \\
\textbf{\textit{O-D-K}}            & \textit{0-OS-0} & \textit{0-OS-1} & \textit{0-OS-1}
                          & \textit{0-OS-0} & \textit{0-OS-0} & \textit{0-OS-1} \\
\noalign{\hrule height 2pt}

& \multicolumn{3}{c!{\vrule width 2pt}}{\textbf{WL5 - T1 profile}}
& \multicolumn{3}{c!{\vrule width 2pt}}{\textbf{WL6 - T1 profile}} \\
\noalign{\hrule height 2pt}
\textbf{\#chiplets}        & 6 & 3 & 3  & 4 & 3 & 2 \\
\textbf{\textit{A-T-S}}    & \textit{64-7-256} x6 & \textit{64-7-256} x3 & \textit{64-7-256} x3
                          & \textit{64-7-256} X4 & \textit{64-7-256} X3 & \textit{64-7-256} x2 \\
\textbf{\textit{I-P-M}}            & \textit{3D-HB-HBM3} & \textit{3D-HB-HBM3} & \textit{2.5D-RDL-3D-HB-HBM3}
                          & \textit{3D-HB-HBM3} & \textit{3D-HB-HBM3} & \textit{3D-HB-HBM3} \\
\textbf{Protocol}          & UCIe 3D & UCIe 3D & UCIe Std-UCIe 3D
                          & UCIe 3D & UCIe 3D & UCIe 3D \\
\textbf{\textit{O-D-K}}            & \textit{0-OS-0} & \textit{0-OS-0} & \textit{0-OS-0}
                          & \textit{0-OS-0} & \textit{0-OS-0} & \textit{0-OS-0} \\
\noalign{\hrule height 2pt}

\end{tabular}%
}
\label{tab:sa_result_compare_table_template1_balance}
\end{table}

Table~\ref{tab:SA-numbers-compare-all} compares the metrics of the generated solution across the three optimization flows, across all optimization templates and all workloads. The table reports the numbers normalized to the solution generated by CarbonPATH. CarbonPATH w/o carbon and ChipletGym occasionally reports lower values for specific metrics, this is largely an artifact of differing modeling assumptions and the prioritization of sustainability.
Specifically, ChipletGym's cost function excludes area constraints and does not penalize high chiplet counts, which lowers its latency. Furthermore, CarbonPATH uses a comprehensive, packaging-aware latency model that yields cycle-accurate latency estimates, in contrast to ChipletGym’s fixed, optimistic D2D latency.



\subsubsection{Comparison of generated solution across all templates}
\label{sec:results-template-optimizations}
Table~\ref{tab:sa_result_compare_table_template1_balance} details the optimized system architecture. Under the T1 optimization template, which assigns equal weight to all the parameters, the final HI system configurations show a consistent trend with ChipletGym yielding a higher chiplet count. Across all workloads, CarbonPATH demonstrates packaging preferences driven by sustainability. For WL2, CarbonPATH favors a 2-chiplet 2.5D integration with \textit{0-OS-1} mapping, whereas ChipletGym and CarbonPATH w/o carbon opt for 3D systems with six and three chiplets, respectively. Similarly, for WL5, CarbonPATH proposes a hybrid 2.5D+3D approach in contrast, both ChipletGym and CarbonPATH w/o carbon remain restricted to 3D architectures with six and three identical chiplets. This confirms that CarbonPATH actively explores the trade-offs between packaging and carbon impact, consistently favoring lower CFP packaging choices over the other two optimization flows. This does come at the cost of latency as shown in Table~\ref{tab:SA-numbers-compare-all}. Since for T1 both CFP and latency are equally weighted. 

\begin{table}[t]
\centering
\caption{Comparison of optimized HI system for template T2.}
\resizebox{\linewidth}{!}{%
\setlength{\arrayrulewidth}{1pt}
\begin{tabular}{!{\vrule width 2pt} l !{\vrule width 2pt} c|c|c !{\vrule width 2pt} c|c|c !{\vrule width 2pt}}
\noalign{\hrule height 2pt}
& \multicolumn{3}{c!{\vrule width 2pt}}{\textbf{WL1 - T2 profile}}
& \multicolumn{3}{c!{\vrule width 2pt}}{\textbf{WL2 - T2 profile}} \\
\noalign{\hrule height 2pt}
\textbf{Parameters} & \textbf{ChipletGym} & \textbf{CarbonPATH wo C} & \textbf{CarbonPATH}
                   & \textbf{ChipletGym} & \textbf{CarbonPATH wo C} & \textbf{CarbonPATH} \\
\hline
\textbf{\#chiplets}         & 3 & 3 & 2   & 6 & 3 & 2     \\
\textbf{\textit{A-T-S}}    & \textit{128-7-1024} X3 & \textit{64-7-256, 96-7-512} X2 & \textit{96-7-512} x2 &
                    \textit{96-7-512} X6 & \textit{96-7-512} x3 & \textit{96-7-512} x2  \\
\textbf{\textit{I-P-M}}           & \textit{3D-HB-HBM3} & \textit{3D-HB-HBM3} & \textit{2.5D-RDL-HBM3} &
                    \textit{3D-HB-HBM3} & \textit{3D-HB-HBM3} & \textit{3D-HB-HBM3} \\
\textbf{Protocol}          & UCIe 3D & UCIe 3D & UCIe Std &
                    UCIe 3D & UCIe 3D & UCIe 3D \\
\textbf{\textit{O-D-K}}           & \textit{0-OS-0} & \textit{0-OS-1} & \textit{0-OS-0} &
                    \textit{0-OS-0} & \textit{0-OS-1} & \textit{0-OS-0} \\
\noalign{\hrule height 2pt}
& \multicolumn{3}{c!{\vrule width 2pt}}{\textbf{WL3 - T2 profile}}
& \multicolumn{3}{c!{\vrule width 2pt}}{\textbf{WL4 - T2 profile}} \\
\noalign{\hrule height 2pt}
\textbf{\#chiplets}          & 3 & 2 & 2   & 6 & 2 & 2  \\
\textbf{\textit{A-T-S}}   & \textit{128-7-1024} X3 & \textit{64-7-256, 96-7-512} & \textit{64-7-256, 96-7-512} &
                    \textit{64-7-256} x6 & \textit{96-7-512} x2  & \textit{64-7-256, 96-7-512}  \\
\textbf{\textit{I-P-M}}           & \textit{3D-HB-HBM3} & \textit{3D-HB-HBM3} & \textit{3D-HB-HBM3}  &
                    \textit{3D-HB-HBM3} & \textit{3D-HB-HBM3} & \textit{3D-HB-HBM3}  \\
\textbf{Protocol}          & UCIe 3D & UCIe 3D & UCIe 3D &
                    UCIe 3D & UCIe 3D & UCIe 3D \\ 
\textbf{\textit{O-D-K}}            & \textit{0-OS-0} & \textit{0-OS-1} & \textit{0-OS-1} &
                    \textit{0-OS-0} & \textit{0-OS-1} & \textit{0-OS-1} \\
\noalign{\hrule height 2pt}
& \multicolumn{3}{c!{\vrule width 2pt}}{\textbf{WL5 - T2 profile}}
& \multicolumn{3}{c!{\vrule width 2pt}}{\textbf{WL6 - T2 profile}} \\
\noalign{\hrule height 2pt}
\textbf{\#chiplets}        & 6 & 2 & 2  & 4 & 2 & 2 \\
\textbf{\textit{A-T-S}}   & \textit{64-7-256} x6 & \textit{64-7-256} x2 &  \textit{64-7-256} x2  &
                    \textit{64-7-256} x4 & \textit{64-7-256} x2 &  \textit{64-7-256} x2 \\
\textbf{\textit{I-P-M}}            & \textit{3D-HB-HBM3} & \textit{3D-HB-HBM3} & \textit{3D-HB-HBM3}  &
                    \textit{3D-HB-HBM3} & \textit{3D-HB-HBM3} & \textit{3D-HB-HBM3}  \\
\textbf{Protocol}          & UCIe 3D & UCIe 3D & UCIe 3D &
                    UCIe 3D & UCIe 3D & UCIe 3D \\
\textbf{\textit{O-D-K}}           & \textit{0-OS-0} & \textit{0-OS-1} & \textit{0-OS-1} &
                    \textit{0-OS-0} & \textit{0-OS-1} & \textit{0-OS-0} \\
\noalign{\hrule height 2pt}
\end{tabular}%
}
\label{tab:sa_result_compare_table_template2_mobile}
\end{table}

Table~\ref{tab:sa_result_compare_table_template2_mobile} details the architectural outcomes for the T2 optimization template. As defined in Table~\ref{tab:optimization_profile_table}, profile T2 prioritizes energy and operational CFP. Under these constraints, CarbonPATH consistently identifies architectures with fewer chiplets than ChipletGym and that match or improve upon the w/o carbon optimization flow, with almost every optimized solution converging to a configuration using 3D hybrid bonding. This preference arises because 3D hybrid bonding offers superior D2D bandwidth, as demonstrated in Sec~\ref{sec:carbon-results}. Since high-bandwidth interconnects minimize data movement energy, they are critical for reducing both energy and operational CFP. In this optimization template T2, the framework consistently yields compact 2-chiplet systems interconnected via high-bandwidth 3D bonding to maximize operational efficiency for most of the workloads. It can be observed that ChipletGym selects large number of chiplets in its proposed HI system solution, which is attributed to two main factors. First, it does not penalize total system area, and second, the differences in the modeling of its D2D communication.

\begin{table}[t]
\centering
\caption{Comparison of optimized HI system for template T3.}
\resizebox{\linewidth}{!}{%
\setlength{\arrayrulewidth}{1pt}
\begin{tabular}{!{\vrule width 2pt} l !{\vrule width 2pt} c|c|c !{\vrule width 2pt} c|c|c !{\vrule width 2pt}}

\noalign{\hrule height 2pt}
& \multicolumn{3}{c!{\vrule width 2pt}}{\textbf{WL1 - T3 profile}}
& \multicolumn{3}{c!{\vrule width 2pt}}{\textbf{WL2 - T3 profile}} \\
\noalign{\hrule height 2pt}

\textbf{Param.} & \textbf{ChipletGym} & \textbf{CarbonPATH wo C} & \textbf{CarbonPATH} &
                      \textbf{ChipletGym} & \textbf{CarbonPATH wo C} & \textbf{CarbonPATH}\\
\hline
\textbf{\#chiplets}          & 3 & 2 & 2  & 6 & 2 & 3 \\
\textbf{\textit{A-T-S}}     & \textit{128-7-1024} X3 & \textit{128-7-1024} X2 & \textit{128-7-1024} x2 &
                                \textit{96-7-512} X6 & \textit{128-7-1024} x2 & \textit{96-7-512} x3  \\
\textbf{\textit{I-P-M}}            & \textit{3D-HB-HBM3} & \textit{2.5D-RDL-HBM3} & \textit{2.5D-RDL-HBM3}  &
                                   \textit{3D-HB-HBM3} & \textit{2.5D-RDL-HBM3} & \textit{2.5D-RDL-HBM3}  \\
\textbf{Protocol}            & UCIe 3D & UCIe Std & UCIe Std &
                               UCIe 3D & UCIe Std & UCIe Std \\
\textbf{\textit{O-D-K}}            & \textit{0-OS-0} & \textit{0-OS-1} & \textit{0-OS-0} &
                                     \textit{0-OS-0} & \textit{0-OS-1} & \textit{0-OS-0} \\
\noalign{\hrule height 2pt}
& \multicolumn{3}{c!{\vrule width 2pt}}{\textbf{WL3 - T3 profile}}
& \multicolumn{3}{c!{\vrule width 2pt}}{\textbf{WL4 - T3 profile}} \\
\noalign{\hrule height 2pt}

\textbf{\#chiplets}        & 3 & 2 & 2    & 2 & 2 & 2      \\
\textbf{\textit{A-T-S}}   & \textit{128-7-1024} X3 & \textit{64-7-256, 128-7-1024}  & \textit{128-7-1024} x2 &
                            \textit{128-7-1024} x2 & \textit{64-7-256, 128-7-1024}  & \textit{64-7-256, 128-7-1024} \\
\textbf{\textit{I-P-M}}   & \textit{3D-HB-HBM3} & \textit{2.5D-RDL-HBM3} & \textit{2.5D-RDL-HBM3} &
                            \textit{3D-HB-HBM3} & \textit{2.5D-RDL-HBM3} & \textit{2.5D-RDL-HBM3} \\
\textbf{Protocol}          & UCIe 3D & UCIe Std & UCIe Std &
                UCIe 3D & UCIe Std & UCIe Std \\
\textbf{\textit{O-D-K}}    & \textit{0-OS-0} & \textit{0-OS-0} & \textit{0-OS-1} &
                             \textit{0-OS-0} & \textit{0-OS-1} & \textit{0-OS-1} \\
\noalign{\hrule height 2pt}

& \multicolumn{3}{c!{\vrule width 2pt}}{\textbf{WL5 - T3 profile}}
& \multicolumn{3}{c!{\vrule width 2pt}}{\textbf{WL6 - T3 profile}} \\
\noalign{\hrule height 2pt}

\textbf{\#chiplets}        & 6 & 2 & 2         & 4 & 3 & 3 \\
\textbf{\textit{A-T-S}}    & \textit{64-7-256} x6 & \textit{96-7-512} x2 & \textit{96-7-512} x2  &
                             \textit{64-7-256} x4 & \textit{64-7-256} x3 & \textit{64-7-256} x3 \\
\textbf{\textit{I-P-M}}   & \textit{3D-HB-HBM3} & \textit{2.5D-RDL-HBM3} & \textit{2.5D-RDL-HBM3} &
                            \textit{3D-HB-HBM3} & \textit{2.5D-RDL-HBM3} & \textit{2.5D-RDL-3D-HB-HBM3}  \\
\textbf{Protocol}          & UCIe 3D & UCIe Std & UCIe Std &
                             UCIe 3D & UCIe Std & UCIe Std-UCIe 3D \\
\textbf{\textit{O-D-K}}    & \textit{0-OS-0} & \textit{0-OS-1} & \textit{0-OS-0} &
                             \textit{0-OS-0} & \textit{0-OS-1} & \textit{0-OS-0} \\
\noalign{\hrule height 2pt}
\end{tabular}%
}
\label{tab:sa_result_compare_table_template3_automotive}
\end{table}

Table~\ref{tab:sa_result_compare_table_template3_automotive} presents the architectural outcomes for the T3 profile, which prioritizes latency and monetary cost. Consistent with these objectives, nearly all workloads converge to designs utilizing RDL-based 2.5D or 2.5D+3D integration and a smaller number of chiplets. This trend is driven by the cost-effectiveness of RDL technology compared to silicon interposers or hybrid bonding.
Because the T3 profile assigns minimal weight to sustainability metrics, the solutions identified by CarbonPATH largely mirror those of the CarbonPATH w/o carbon flow. WL6 presents a notable exception. In this case, CarbonPATH identifies a distinct 2.5D+3D configuration. Compared to the 3-chiplet 3D solution found by the baseline, this hybrid architecture offers comparable latency and lower monetary cost, with the added co-benefit of reduced embodied CFP. In contrast, in ChipletGym, we observe 4 chiplets in 3D systems, as it prioritizes latency over area and cost. A distinction in WL2 is that CarbonPATH proposes a three-chiplet configuration, whereas the w/o Carbon version selects only two. Although CarbonPATH uses more chiplets, the smaller individual chiplet size yields a system that is both cost-effective and more sustainable than the larger two-chiplet alternative.

Table~\ref{tab:sa_result_compare_table_template4_wearables} presents the architectural outcomes for profile T4, which jointly optimizes energy, area, operational CFP, and embodied CFP. Compared to the baselines, CarbonPATH reveals a significant divergence in the optimized architectures. The results display a high degree of workload dependency: while some workloads retain hybrid bonding for its efficiency, others shift toward RDL or even monolithic designs (as seen in WL3). This confirms that incorporating strict sustainability weights forces a move away from complex, high-emission packaging technologies. The framework effectively trades off packaging sophistication for carbon savings, selecting mature technologies unless the workload explicitly requires the bandwidth of advanced integration.

\begin{table}[t]
\centering
\caption{Comparison of optimized HI system for template T4.}
\resizebox{\linewidth}{!}{%
\setlength{\arrayrulewidth}{1pt}
\begin{tabular}{!{\vrule width 2pt} l !{\vrule width 2pt} c|c|c !{\vrule width 2pt} c|c|c !{\vrule width 2pt}}
\noalign{\hrule height 2pt}
& \multicolumn{3}{c!{\vrule width 2pt}}{\textbf{WL1 - T4 profile}}
& \multicolumn{3}{c!{\vrule width 2pt}}{\textbf{WL2 - T4 profile}} \\
\noalign{\hrule height 2pt}
\textbf{Param} & \textbf{ChipletGym} & \textbf{CarbonPATH wo C} & \textbf{CarbonPATH} 
                    & \textbf{ChipletGym} & \textbf{CarbonPATH wo C} & \textbf{CarbonPATH}\\
\hline
\textbf{\#chiplets}        & 3 & 4 & 3 & 6 & 2 & 2        \\
\textbf{\textit{A-T-S}}   & \textit{128-7-1024} X3   & \textit{64-7-256, 96-7-512} X3 & \textit{64-7-256} x3 &
                            \textit{96-7-512} X6     & \textit{96-7-512} x2           & \textit{64-7-256} x2  \\
\textbf{\textit{I-P-M}}   & \textit{3D-HB-HBM3} & \textit{3D-HB-HBM3} & \textit{2.5D-RDL-3D-HB-HBM3} &
                            \textit{3D-HB-HBM3} & \textit{3D-HB-HBM3} & \textit{3D-HB-HBM3} \\
\textbf{Protocol}          & UCIe 3D & UCIe 3D & UCIe Std-UCIe 3D &
                        UCIe 3D & UCIe 3D & UCIe 3D \\
\textbf{\textit{O-D-K}}   & \textit{0-OS-0} & \textit{0-OS-1} & \textit{0-OS-0} &
                            \textit{0-OS-0} & \textit{0-OS-1} & \textit{0-OS-0} \\
\noalign{\hrule height 2pt}
& \multicolumn{3}{c!{\vrule width 2pt}}{\textbf{WL3 - T4 profile}}
& \multicolumn{3}{c!{\vrule width 2pt}}{\textbf{WL4 - T4 profile}} \\
\noalign{\hrule height 2pt}

\textbf{\#chiplets}         & 3 & 5 & 1 & 2 & 3 & 2 \\
\textbf{\textit{A-T-S}}   & \textit{128-7-1024} X3 & \textit{64-7-256} x5  & \textit{128-7-1024} &
                            \textit{128-7-1024} x2 & \textit{64-7-256} x3  & \textit{64-7-256} x2  \\
\textbf{\textit{I-P-M}}           & \textit{3D-HB-HBM3} & \textit{3D-HB-HBM3} & \textit{2D-NA-DDR5} &
                                    \textit{3D-HB-HBM3} & \textit{3D-HB-HBM3} & \textit{2.5D-RDL-HBM3} \\
\textbf{Protocol}          & UCIe 3D & UCIe 3D & NA &
                    UCIe 3D & UCIe 3D & UCIe Std \\
\textbf{\textit{O-D-K}}            & \textit{0-OS-0} & \textit{0-OS-0} & \textit{1-WS-0} &
                                     \textit{0-OS-0} & \textit{0-OS-1} & \textit{0-OS-0} \\
\noalign{\hrule height 2pt}
& \multicolumn{3}{c!{\vrule width 2pt}}{\textbf{WL5 - T4 profile}}
& \multicolumn{3}{c!{\vrule width 2pt}}{\textbf{WL6 - T4 profile}} \\
\noalign{\hrule height 2pt}

\textbf{\#chiplets}        & 6 & 4 & 2  & 4 & 2 & 2 \\
\textbf{\textit{A-T-S}}   & \textit{64-7-256} x6 & \textit{64-7-256} x4 & \textit{64-7-256} x2  &
                            \textit{64-7-256} x4 & \textit{64-7-256} x2 & \textit{64-7-256} x2 \\
\textbf{\textit{I-P-M}}               & \textit{3D-HB-HBM3} & \textit{3D-HB-HBM3} & \textit{3D-HB-HBM3} &
                                        \textit{3D-HB-HBM3} & \textit{3D-HB-HBM3} & \textit{3D-HB-HBM3} \\
\textbf{Protocol}          & UCIe 3D & UCIe 3D & UCIe 3D &
                    UCIe 3D & UCIe 3D & UCIe 3D \\
\textbf{\textit{O-D-K}}   & \textit{0-OS-0} & \textit{0-OS-0} & \textit{0-OS-0} &
                            \textit{0-OS-0} & \textit{0-OS-1} & \textit{0-OS-0} \\
\noalign{\hrule height 2pt}
\end{tabular}%
}
\label{tab:sa_result_compare_table_template4_wearables}
\end{table}

\subsubsection{Simulated annealing runtimes}
\noindent
Table~\ref{tab:run-time-table} reports the runtime of CarbonPATH for different workloads under the T1 optimization template. Across all workloads, the SA engine converges in under two hours on an AMD EPYC 7313 16-core processor. The most time-intensive component is the cost function evaluation, which invokes ScaleSim. However, with our simulation cache implemented (Sec~\ref{sec:sa-runtime-considerations}), we can still ensure reasonable runtimes for CarbonPATH. For example, without caching, WL5 requires 363 minutes to converge, whereas with the simulation cache, convergence time drops to just 73 minutes.  

\begin{table}[t]
\centering
\caption{CarbonPATH runtimes across for T1 template.}
\footnotesize 
\begin{tabular}{!{\vrule width 2pt}c|c!{\vrule width 2pt}c|c!{\vrule width 2pt}}
\noalign{\hrule height 2pt} 
\textbf{Workload} & \textbf{Runtime} & \textbf{Workload} & \textbf{Runtime} \\ 
\noalign{\hrule height 2pt} 

1 & 61 min & 4 & 64 min\\ \hline
2 & 109 min & 5 & 73 min\\ \hline
3 & 65 min & 6 & 57 min \\ 


\noalign{\hrule height 2pt} 
\end{tabular}
\label{tab:run-time-table}
\end{table}

\balance
\section{Validation of CarbonPATH}
\label{sec:validation}

The accuracy of CarbonPATH depends on the fidelity of its underlying models. To ensure reliability, we derive parameter values from trusted sources and direct synthesis. Chiplet area and power are obtained by synthesizing Verilog implementations of various systolic array sizes using Synopsys Design Compiler with the ASAP7 PDK. Results for other technology nodes are scaled following established methodologies~\cite{eco-chip}, assuming a 12.5\% activity factor for power analysis. SRAM energy values are drawn from~\cite{seo-sram-area-power}, while DRAM energy consumption is estimated from~\cite{dram-energy-1,dram-energy-2}.  Latency is modeled using the cycle-accurate simulator ScaleSim~\cite{scalesim}, evaluated at 1~GHz for 7~nm, consistent with synthesis, and frequency-scaled for other nodes following~\cite{freq-scaling}. We use ECO-CHIP~\cite{eco-chip} to estimate embodied and operational CFP. This allows CarbonPATH to capture realistic metrics under different workload mappings and technology assumptions. Die-to-die energy per bit is taken from published efficiency data for BoW, UCIe, and AIB~\cite{bow-pjb,ucie-pjb,aib-pjb}. Finally, dollar and memory costs are modeled using wafer cost estimates from~\cite{cost-per-wafer,mem_cost}.  

By grounding the framework in a combination of direct synthesis, cycle-accurate simulation, and trusted reference data, CarbonPATH yields estimates consistent with established theoretical principles for heterogeneous integration. This validation gives confidence that our analyses not only reflect realistic system behavior but also offer meaningful insights into the tradeoffs between PPAC and CFP.  
CarbonPATH is being released as an open-source framework to encourage broad adoption~\cite{carbonpath-gh}. It has been designed to be plug-and-play friendly: if models need to be modified or updated, users can simply change the input values or references to reflect their own technology assumptions (e.g., yield, defect density, or proprietary efficiency data). While detailed physical design and signoff flows remain essential for final implementation and validation, our models enable designers to explore a large and complex design space supporting early-phase pathfinding. Crucially, at this stage of analysis, the model's fidelity in capturing relative trends matters more than any single absolute point estimate, which is why we present normalized results throughout. Normalization reduces the extent to which the results depend on any single technology assumption and instead highlights relative differences across architectures, packaging options, and workload mappings. The primary value of CarbonPATH is not in claiming an exact PPAC or CFP value for a given design, but rather in providing a framework for consistent comparisons across configurations, helping designers determine which system is better and under what conditions.

\section{Conclusion}
\label{sec:conclusion}
We presented CarbonPATH, a pathfinding framework for early-stage co-design of HI systems targeting AI accelerators. CarbonPATH explores a broad design space spanning workload mapping, chip architecture, chiplets, sustainability, and packaging, while evaluating system-level tradeoffs in PPAC. By developing analytical models for compute, memory, and die-to-die communication, CFP, and using a cycle-accurate simulator, the framework captures the impact of packaging technology, interconnect topology, and workload partitioning on system PPAC.  We also compared CarbonPATH with ChipletGym, a recent framework for optimizing chiplet-based systems. CarbonPATH finds more optimized solutions by searching a larger design space and by employing more detailed modeling of latency, energy, area, cost, and CFP. CarbonPATH is open-source and available at~\cite{carbonpath-gh}.

\newpage
\bibliographystyle{misc/ieeetr2}
\bibliography{misc/bibfile}

\end{document}